\documentclass[twocolumn,showkeys,preprintnumbers]{revtex4}
\usepackage{graphicx}
\usepackage{epsfig}
\newcommand{\GC} {\ifmmode gamma-catcher \else $\gamma$-catcher \fi}

\begin{document}
\title{A Comparison of the Performance of Compact Neutrino Detector Designs
for Nuclear Reactor Safeguards and Monitoring}
\author{R.~W.~McKeown}
\email{rwm33@drexel.edu} 
\affiliation{Physics Dept., Drexel University, 3141 Chestnut St.,
Philadelphia, PA 19104}
\author{D.~E.~Reyna}
\email{reyna@anl.gov}
\affiliation{HEP Division, Argonne National Laboratory, 9700 S. Cass Ave.,
Argonne, IL 60439}
\date{October 27, 2006}

\begin{abstract}
There has been an increasing interest in the monitoring of nuclear fuel
for power reactors by detecting the anti-neutrinos produced during operation.
Small liquid scintillator detectors have already demonstrated sensitivity
to operational power levels, but more sensitive monitoring requires 
improvements in the efficiency and uniformity of these detectors.  In this
work, we use a montecarlo simulation to investigate the detector 
performance of 
four different detector configurations.  Based on the analysis of
neutron detection efficiency and positron energy response, we find that
the optimal detector design will depend on the goals and restrictions
of the specific installation or application.  We have attempted to 
present the relevant information so that future detector development can
proceed in a profitable direction. 
\end{abstract}



\maketitle{}

\section{Introduction}
Recently, there has been an increasing interest in the
monitoring and safeguarding of nuclear power reactors.  
Over the last few years,
a group from Livermore and Sandia National Laboratories has been 
demonstrating the feasibility of a small simple detector to monitor the
anti-neutrino production during the operation of a nuclear power reactor
located in San Onofre, California\cite{Bernstein:2001cz}.  
By using the anti-neutrinos
which are produced by the uranium and plutonium fuel itself during the
fission process, the reactor fuel can be continuously monitored in a
non-invasive way.  The ability to
monitor the nuclear fuel composition in real-time has advantages for
both limiting the proliferation of nuclear material as well as increasing
the operational efficiency of power generation\cite{r-mon,r-mon2}.  

The current experience at San Onofre has demonstrated that a small 
neutrino detector located within 25~m of the reactor core is easily 
sensitive to the power level at which the nuclear reactor is being operated.
However, more sensitive tests to determine the fuel burn-up and fuel
composition, while showing promising results, have shown the limitations 
of the detector design.  

This work is an attempt to evaluate several possible directions for improved
compact neutrino detector designs using the latest simulations available
to the reactor neutrino community.  The detectors are based on the 
conventional technology of liquid scintillator and photomultiplier tubes.
The anti-neutrino event signature is the inverse beta-decay process:
$\bar{\nu}_e + p \rightarrow e^+ + n$.  This yields a coincident event
signature, from the prompt positron annihilation and the delayed neutron
capture, which is relatively free from background contamination.  
The use of a liquid scintillator that is doped with gadolinium 
improves the signal
to background further by reducing the neutron capture time and increasing
the energy which that capture releases.  

The simulations were based on
the open-source libraries of the {\em Generic Liquid-scintillator anti-neutrino
detector Geant4 simulation} (GLG4sim)\cite{glg4sim,geant4} 
which have been extensively
used in the KamLAND experiment\cite{Araki:2004mb}.  In addition, we have 
made 
use of several of the improvements that have been developed within 
the Double Chooz collaboration\cite{Ardellier:2006mn}.  Specifically, the
development of high quality Gd doped liquid scintillator has been carefully
studied and implemented within the simulation, as well as detailed 
improvements in material properties and PMT performance.

For this study, it was considered that there are two major features of the
detector response which will be important for reactor monitoring.  The first
concerns the efficiency for detecting events in which the neutron is 
captured on Gd within the
fiducial volume of the detector.  The overall acceptance for anti-neutrino 
events is a product of positron and neutron identifications.  However, the 
positron detection efficiency is largely a function of the environmental 
conditions and can more directly be improved by shielding the detector to 
allow a reduction in the minimum energy threshold for acceptance. 
The neutron efficiency, on the other hand, can be improved through design 
features of the detector---ensuring that the photons released through 
the n-Gd capture are all detected within the scintillator to allow good 
separation between neutron-like and positron-like energy depositions. 

The second detector performance characteristic studied
was the uniformity of energy response for positron events.  If  
sufficient statistical power is available, the positron energy 
spectrum---which is directly related to the parent neutrino spectrum---will
contain all needed information to extract the specific fuel compositions
in real-time.  This, however, requires that the detector itself does not
contribute any meaningful distortion to the energy spectrum, whether it
be a non-linear response or a non-uniformity based on position dependence.

\section{Detector Designs}
The detector designs that were studied were intended to explore various
methods of reducing the overall size of a detector while maintaining 
sufficient statistical acceptance to be relevant for the desired monitoring.
The baseline fiducial volume of 2~m$^3$ ($\sim$2 tons) was chosen for 
all designs to provide a sufficient event rate for any detector
within $\sim$60~m of a reactor core.  In addition, an attempt was made
to keep the overall size of the detector small---all dimensions less than
3--4~m.  

The four designs, described in detail below, make use of the latest
technological developments which have been implemented within
the reactor neutrino community.  All of the designs include a fiducial 
``target'' region which is filled with a Gd doped liquid scintillator. 
Some of the designs include a ``gamma-catcher'' which is an un-doped 
liquid scintillator region surrounding the target whose purpose is to 
completely absorb the photons released by the n-Gd capture.  Acrylic is
used wherever a volume boundary requires optical transparency.
Also, in all designs,
8'' PMTs have been used and are installed within a 1~m mineral oil buffer 
whose purpose is to reduce the singles background by attenuating the 
radioactive photons emitted by the $^{40}$K in the PMT glass.

\subsection*{Design~1: Miniature Physics Design}
\begin{figure}[htb]
\begin{center}
\includegraphics[height=0.75\columnwidth]{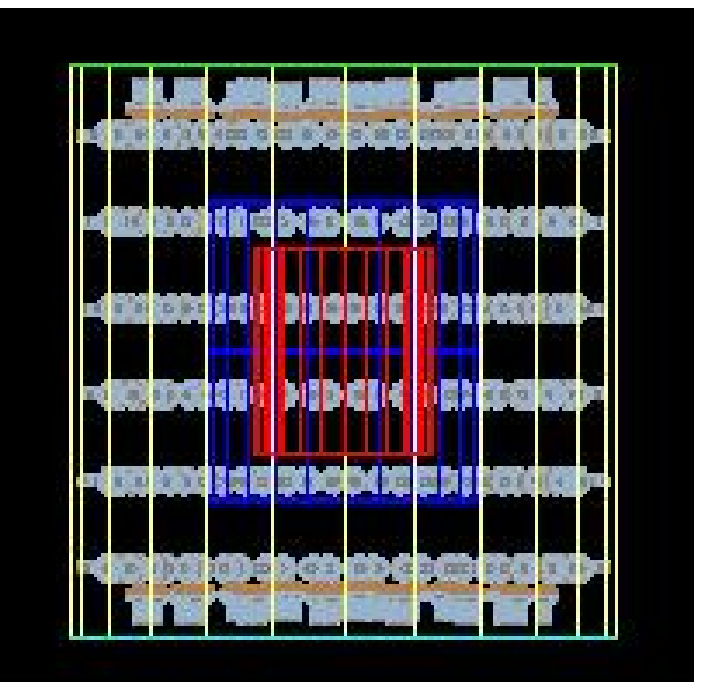}
\end{center}
\label{fig:sdesign}
\end{figure}
This design is a scaled down version, based on the state-of-the-art 
detectors that have been developed for the next generation reactor based 
neutrino oscillation measurement at Double Chooz.  The detector uses 
nested acrylic cylinders to contain the target and gamma-catcher volumes.
The 2 m$^3$ target has a height of 1.5~m and diameter 1.3~m while the 
gamma-catcher has a uniform thickness of 35~cm.  The 1~m thick mineral oil 
buffer completely surrounds the active scintillator regions and is contained
within a stainless steel vessel.  The stainless steel is considered to be
polished with a reflectivity of $\sim$40\%.  A total of 282 PMTs are installed
on the inner wall of the stainless steel vessel for an active coverage of
$\sim$15\%.  

The total dimension of this detector design is 4.2~m in height and 4~m in 
diameter.  While this is a little larger than the stated goal, it is 
expected that this design will provide the optimal detector performance
and is therefore a useful benchmark against which the other detector
designs can be compared.

\subsection*{Design~2: Two Sided Cylinder Design}
\begin{figure}[htb]
\begin{center}
\includegraphics[height=0.75\columnwidth]{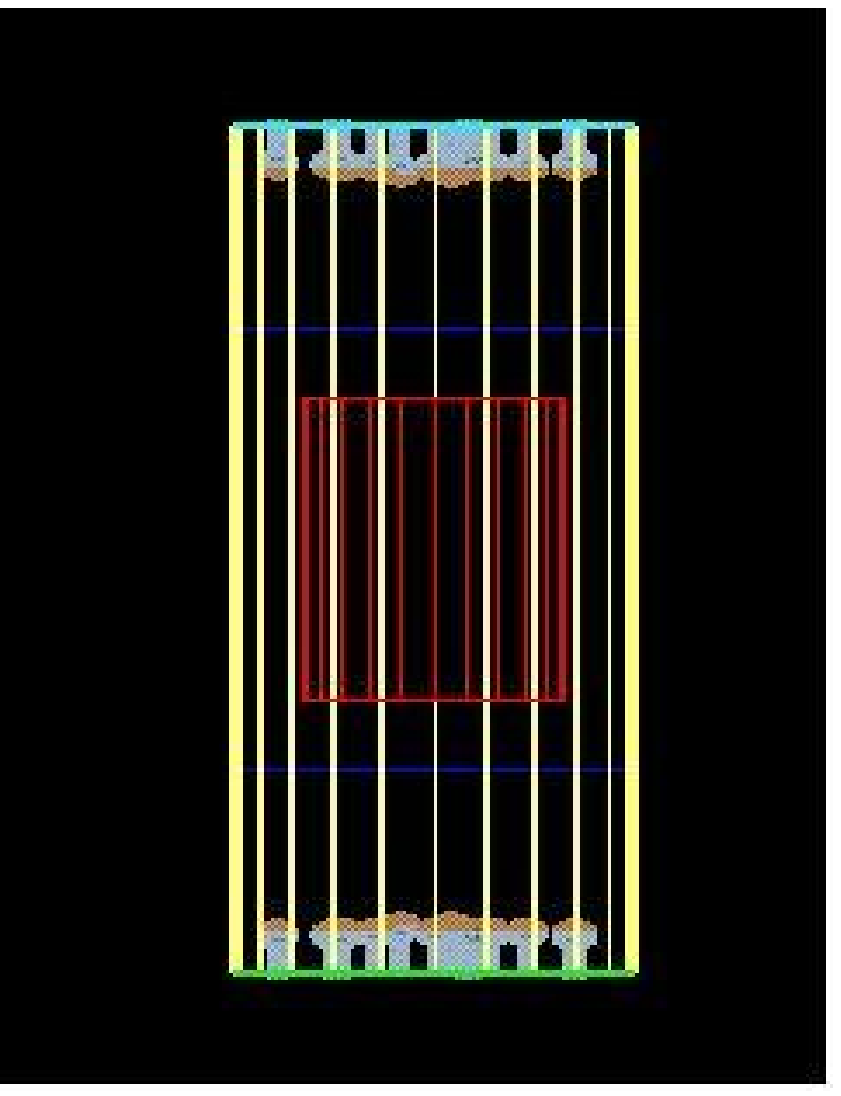}
\end{center}
\label{fig:cdesign}
\end{figure}
This design attempts to reduce the overall footprint and cost of the previous 
design by using reflective walls to eliminate most of the PMT coverage
and the associated mineral oil volume.  Identical to Design~1, this
design has a 2 m$^3$ target cylinder of 1.5~m height and 1.3~m diameter and
a uniform 35~cm gamma-catcher.  However, in this design, PMTs are only located
at the top and bottom of the detector.  Therefore, the outer vessel matches 
the diameter of the gamma-catcher, but extends
an extra meter at the top and bottom to house the mineral oil buffer
and the PMTs.  The barrel region of the outer vessel is defined to
have a diffuse reflective surface---approximating Tyvek---with 
reflectivity of $\sim$90\%, while the top and bottom are defined to 
be non-reflective.  A total of 30 PMTs (15 each at the top and bottom)
provide an effective covering of $\sim$15\%.  The overall dimension of 
this detector is 4.2~m in height and 2~m in diameter.

\subsection*{Design~3: Two Sided Minimal Design}
\begin{figure}[htb]
\begin{center}
\includegraphics[height=0.75\columnwidth]{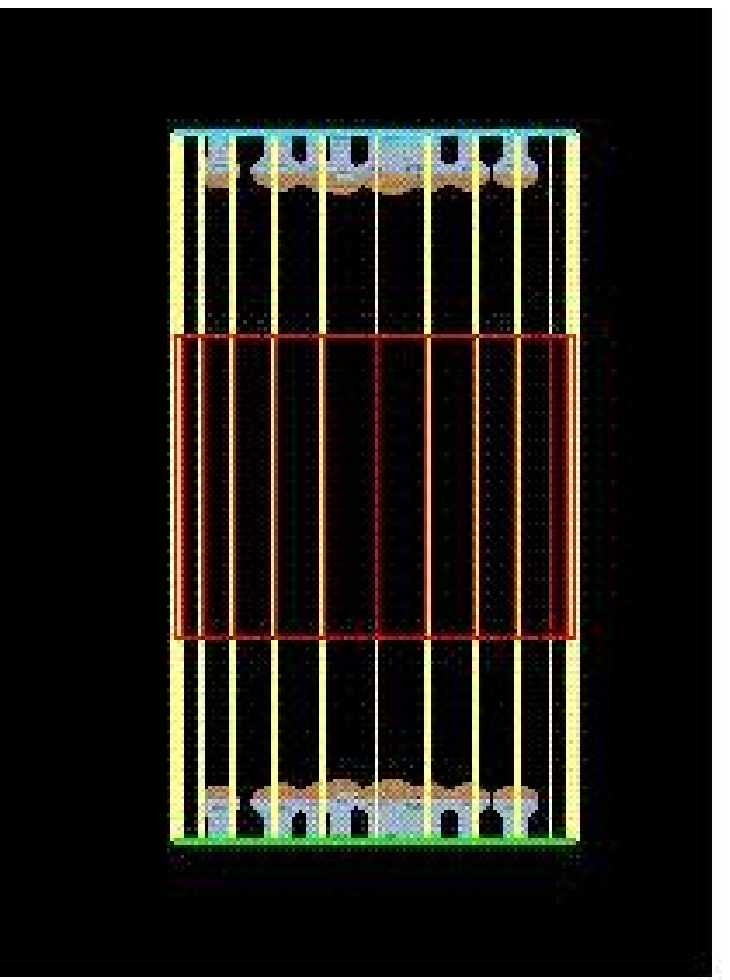}
\end{center}
\label{fig:cyl2design}
\end{figure}
In an attempt to reduce further the footprint of Design~2, this detector
has eliminated the gamma-catcher entirely. To compensate for the expected
efficiency loss at the edge of the target, the target volume has been 
expanded to match the diameter of the gamma-catcher in Designs 1 and 2.
Thus the target has dimensions of 1.5~m height and 2~m in diameter for 
a total volume of 4.71 m$^3$---more than double the previous values.
This provides a good test of the relative benefit of increased efficiency 
vs. increased fiducial volume.  Similar to Design~2, a 1~m mineral oil 
buffer and PMTs
are placed at the top and bottom of the detector.  As in Design~2, the
barrel region of the outer vessel has a diffuse reflective surface 
consistent with Tyvek (90\% reflectivity) and the top and bottom 
of the outer vessel are non-reflective.  A total of
30 PMTs (15 each at the top and bottom) provide $\sim$15\% coverage.
The total dimension of this detector is 3.5~m in height and 2.0~m in diameter.

\subsection*{Design~4: Single Sided Design}
\begin{figure}[!h]
\begin{center}
\includegraphics[height=0.75\columnwidth]{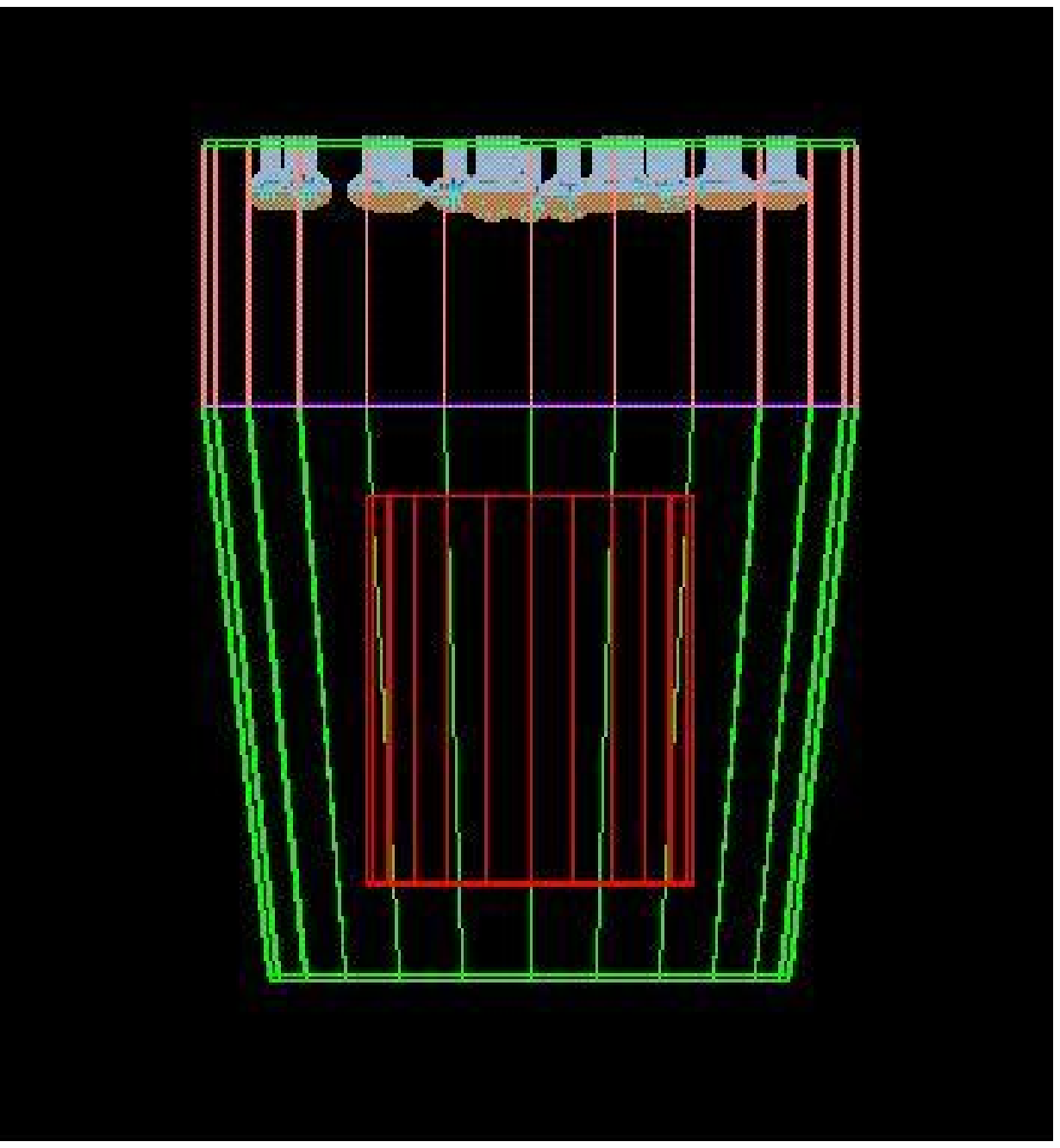}
\end{center}
\label{fig:design}
\end{figure}
This detector is an alternate attempt to reduce the footprint of the 
detector, while maintaining the use of a gamma-catcher.  In this design,
PMTs are only used at the top of the detector.  The target region has
the same 2 m$^3$ volume as Designs 1 and 2---1.5~m height and 1.3~m diameter.
In order to improve light collection at the single end, the gamma-catcher 
is now a conic section with a base diameter of 2~m, 
a height of 2.2~m, and a top diameter of 2.50~m.  This provides a thickness
of 35~cm or greater around the target region.  With a diffuse reflective
surface similar to Tyvek (reflectivity $\sim$90\%) the chosen 6.5$^\circ$
angle of the outer vessel ensures that light from anywhere in
the target volume will have a path length of less than 14~m to the plane of the
PMTs.  While the bottom of the detector is now also reflective, the
top is still defined to be non-reflective.  A total of 24 PMTs are installed
at the top of the 1~m mineral oil buffer providing $\sim$15\% active converage.
This design has a total dimension of 3.2~m in height and 2.5~m in diameter.

\section{Neutron Identification Efficiency}
The identification of an inverse beta-decay event is largely dependent on
the ability to identify the signature of the neutron capture by gadolinium.
The n-Gd capture process releases between 3--10 photons with a total energy
of $\sim$8~MeV.  Since the positron energy deposition is usually between
1--6~MeV, it is common to define a neutron capture signature as any
energy deposition of 6--10~MeV.  In addition to inefficiencies which are
caused by losses of some of the photons from the n-Gd capture, there will 
also be some neutrons which will instead be captured on free protons in
the scintillator.  This n-proton capture releases a single photon of
energy 2.2~MeV and would not satisfy the selection criteria.

For this study, we generated a uniform distribution of neutrons within the
target volume.  The neutrons were generated with a kinetic energy of
2.5~MeV in a random direction.  This kinetic energy is higher than that 
expected from inverse beta-decay events, but subsequent cross-checks 
demonstrated that the results presented here are consistent with those of
lower energy neutrons.  The simulation libraries described above were used to
montecarlo the neutron thermalization and eventual capture.  The ensuing 
optical photons from the scintillator were tracked to the photocathodes of
the PMTs.  Within the simulation, the PMTs used a radially dependent 
quantum efficiency which is consistent with experimental testing.  This 
yielded a photon hit-count per PMT that is roughly equivalent to a number
of photoelectrons.  This photon hit-count has been used for all comparisons
of detector response---avoiding the complications of varying electronics
and data acquisition systems.

\begin{figure}[h]
\begin{center}
\includegraphics[width=\columnwidth]{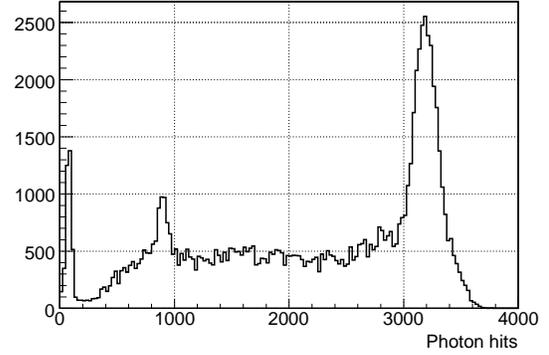}
\end{center}
\caption{Total detected energy deposition for neutrons captured within the 
target volume of detector Design~1.  The photon hits represent the simulated
number of photoelectrons at the PMTs.  The peak at around 3200 photon hits
represents the 8~MeV deposition due to n-Gd capture while the peak at around
900 photon hits represents the 2.2~MeV deposition due to n-p capture.}
\label{F-nCap1}
\end{figure}

By using the truth information from the montecarlo
simulation, events were selected in which the neutron was stopped or captured
within the target volume.  All PMT hits recorded within 100ns were summed
together to create an effective total charge deposition. Fig.~\ref{F-nCap1}
shows the photon hit spectrum for these events from Design~1.
One can quite clearly see the n-Gd capture peak at around 3200 photon 
hits (representing $\sim$8~MeV) and the n-p capture peak at around 900
photon hits (representing 2.2~MeV).  

\begin{figure*}[hbt]
\includegraphics[width=\columnwidth]{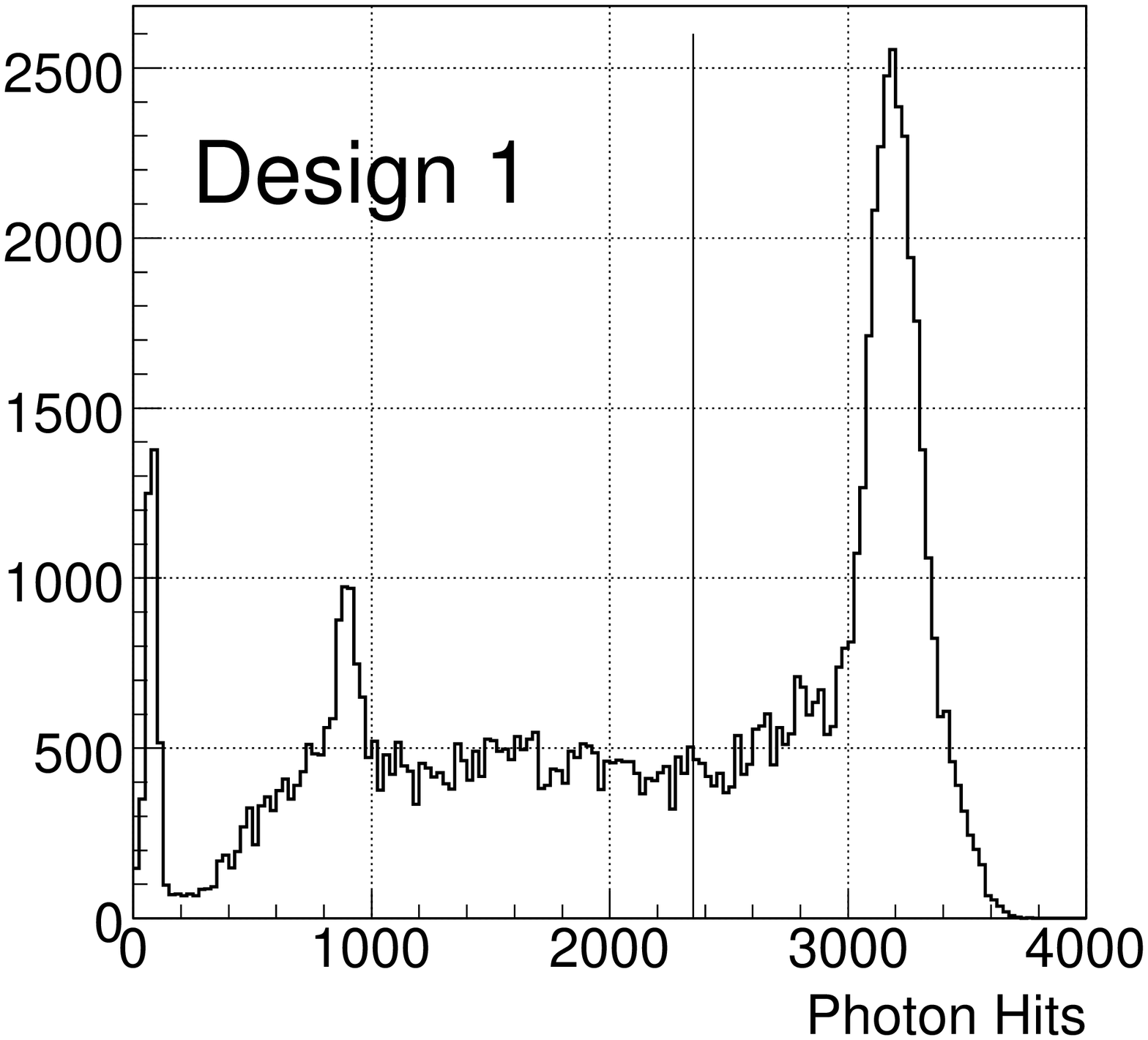}
\hfill
\includegraphics[width=\columnwidth]{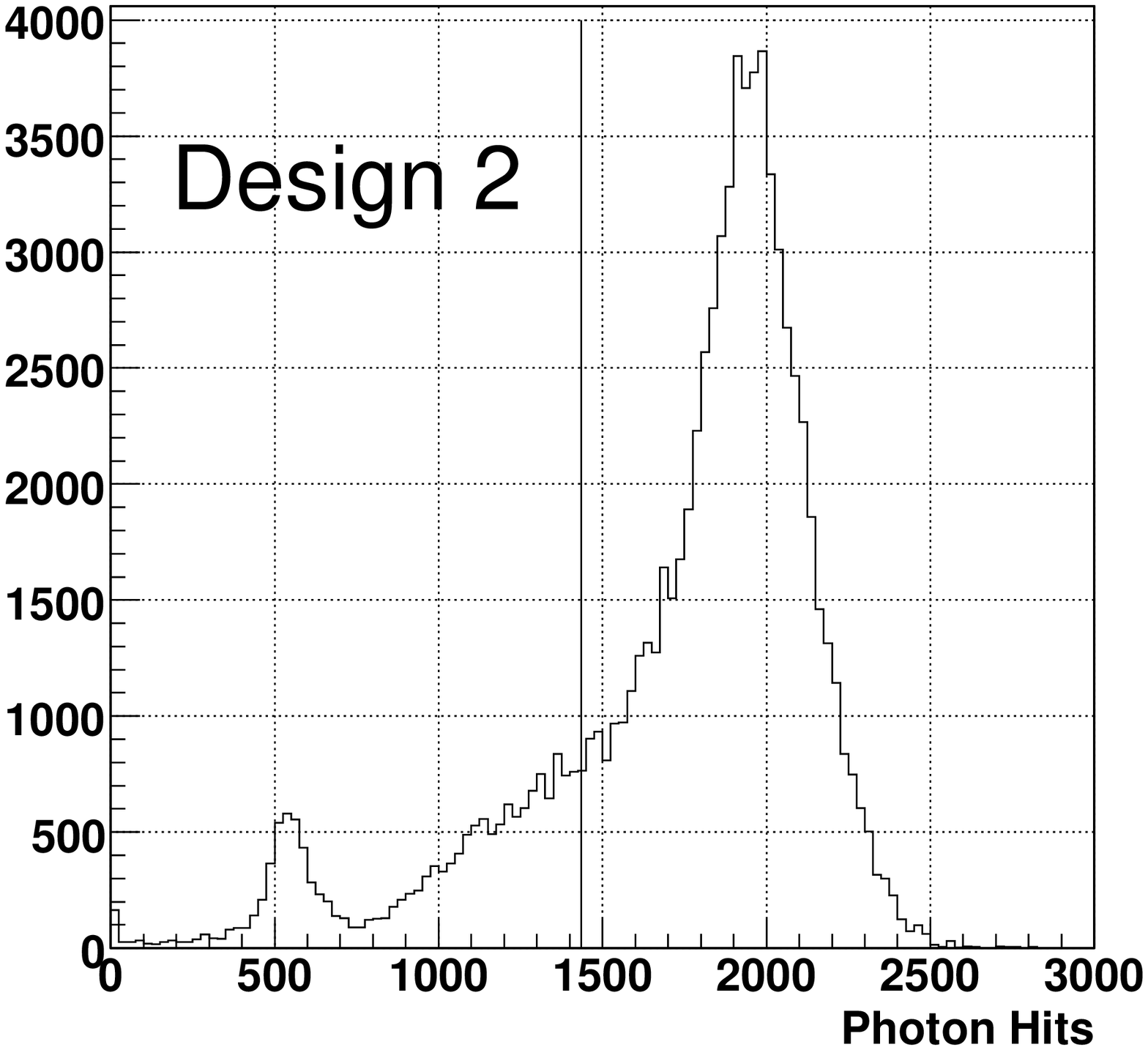}

\includegraphics[width=\columnwidth]{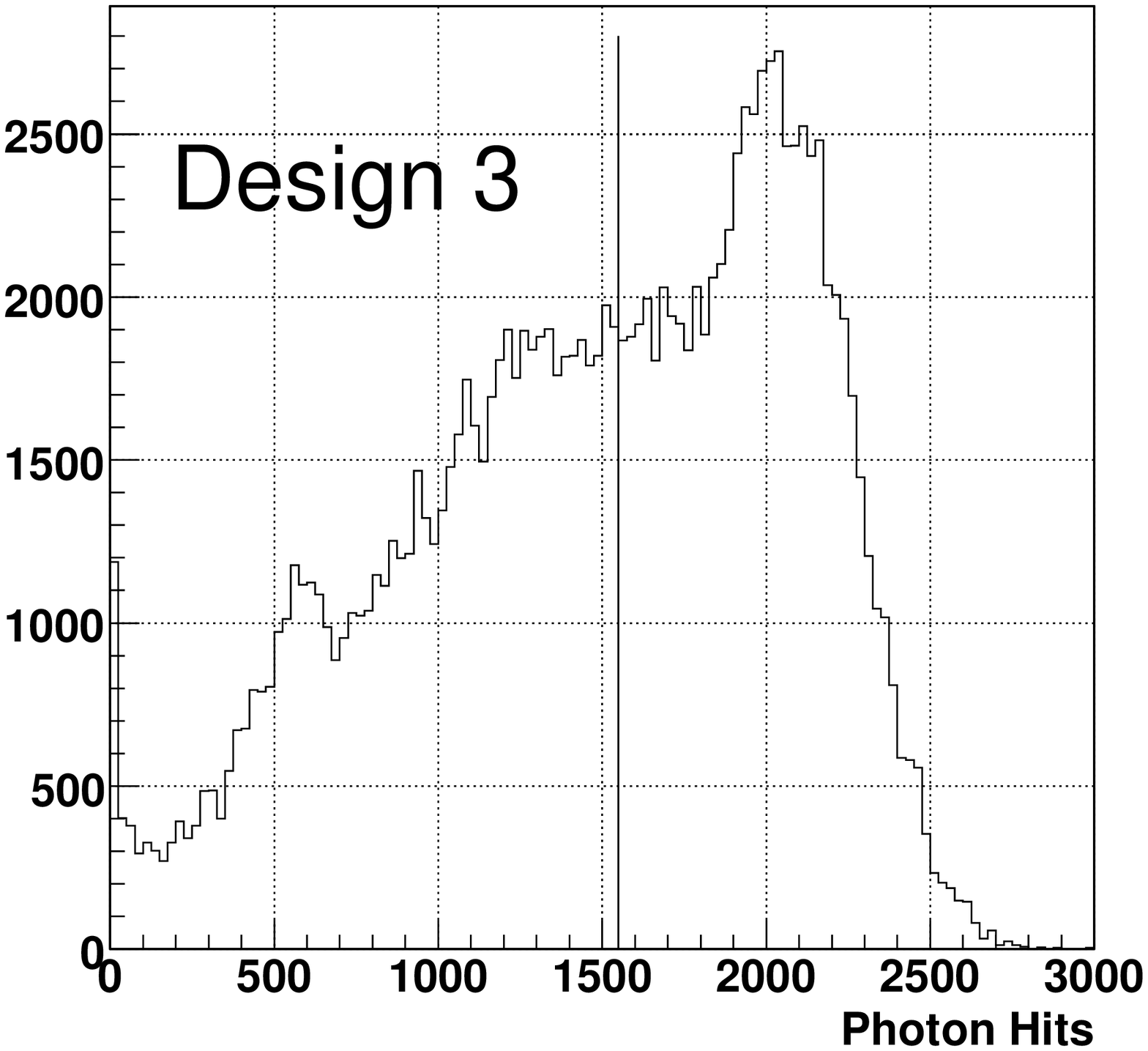}
\hfill
\includegraphics[width=\columnwidth]{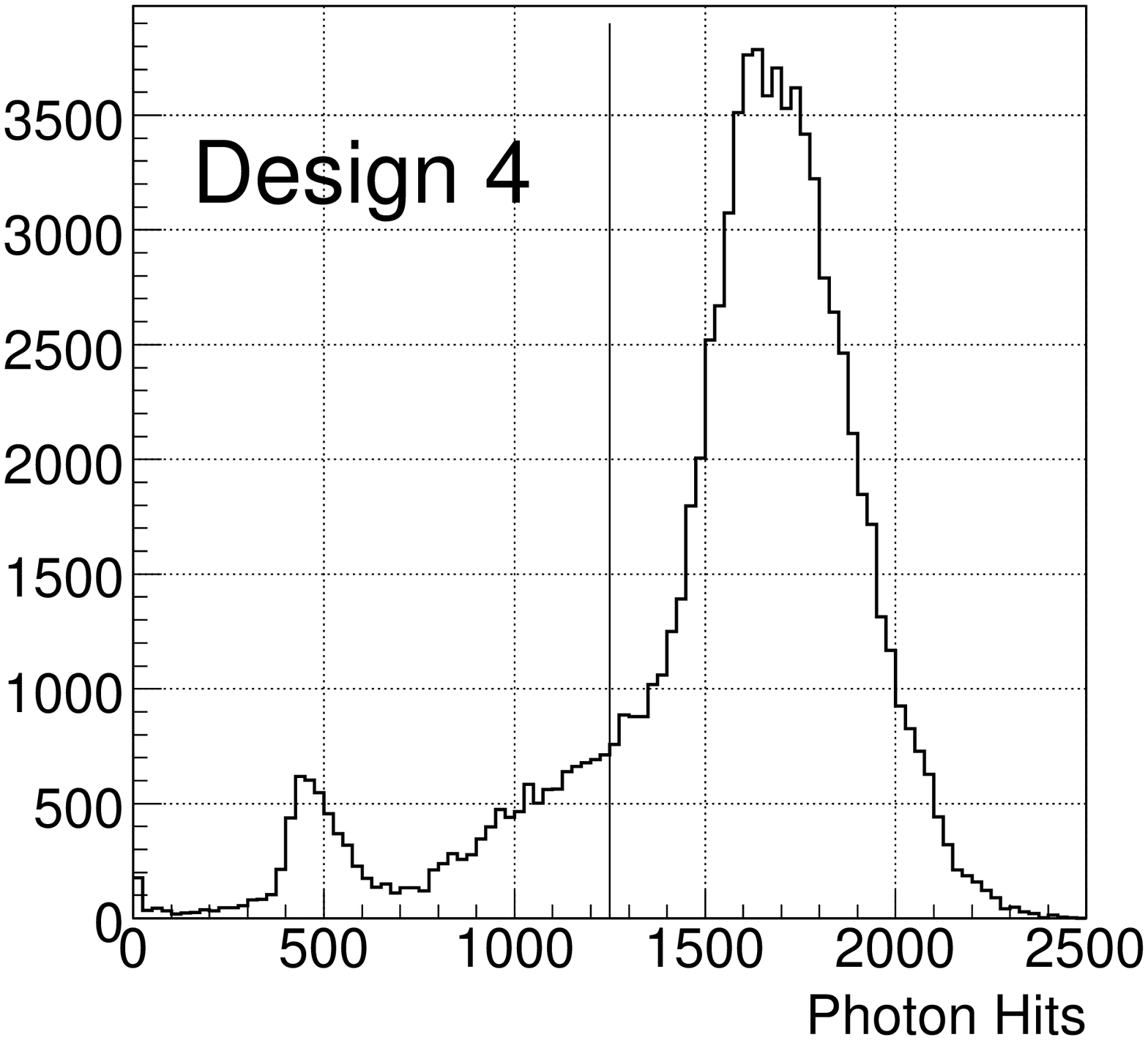}
\caption{Total detected energy deposition for neutrons captured within
the target volume for each detector design, as labeled.  The energy 
is shown in units of photon hits which are a montecarlo representation of
photoelectrons from the PMTs.  Also shown, for each detector design, 
is the threshold for selecting an event in which the neutron was captured
on gadolinium.  The threshold was chosen as the point $\frac{2}{3}$ 
between the fitted peak values of the n-Gd and n-p captures.}
\label{F-nCapCuts}
\end{figure*}

To avoid uncertainties in the quality of calibrations or the variations
in the linearity of response for the various detector designs, it was decided
that the minimum threshold for n-Gd capture identification would be 
defined as the number of photon hits located at the point $\frac{2}{3}$ 
between the fitted peak values of the n-Gd and n-p captures.  This value
would be roughly equivalent to $\sim$6~MeV and provided
a very robust and uniform definition to be applied to all four designs.
In Fig.~\ref{F-nCapCuts} the photon hit spectrum for each of the four 
detector designs is shown with the applicable threshold value superimposed.

With the application of these thresholds, the overall neutron detection 
efficiency for each design is shown in Table~\ref{T-eff}.
It was an initial surprise that Design~1 had such a low efficiency.  After
further investigation, it was understood that the small dimension of
the target and gamma-catcher allow many of the photons from the n-Gd
capture to penetrate to the inactive mineral oil buffer before depositing their
energy.  In the other three designs, these photons are being absorbed 
in the outer wall of the vessel and some of that energy is re-emitted back into
the active scintillator volumes.  This quite clearly demonstrates that
while Design~1 is optimized for larger detectors with fiducial volume
of greater than 10~m$^3$, it is not necessarily optimal for a more 
compact design.
\begin{table}[htb]
\centering
\caption{Total efficiency for selection of neutron captures for each 
detector design.  The sample is based on events in which a neutron stopped 
within the target volume.  The neutron capture is identified by a number
of photon hits greater than the threshold shown in Fig.~\ref{F-nCapCuts}.}
\label{T-eff}
\begin{tabular}{l c}
\hline
Detector & Efficiency \\
\hline \hline
Design~1 &  51.6\% \\ 
Design~2 &  80.4\% \\
Design~3 &  50.2\% \\
Design~4 &  83.4\% \\
\hline
\end{tabular}
\end{table}

\subsection*{Uniformity of Neutron Identification}
\begin{figure*}[htbp]

\includegraphics[width=\columnwidth]{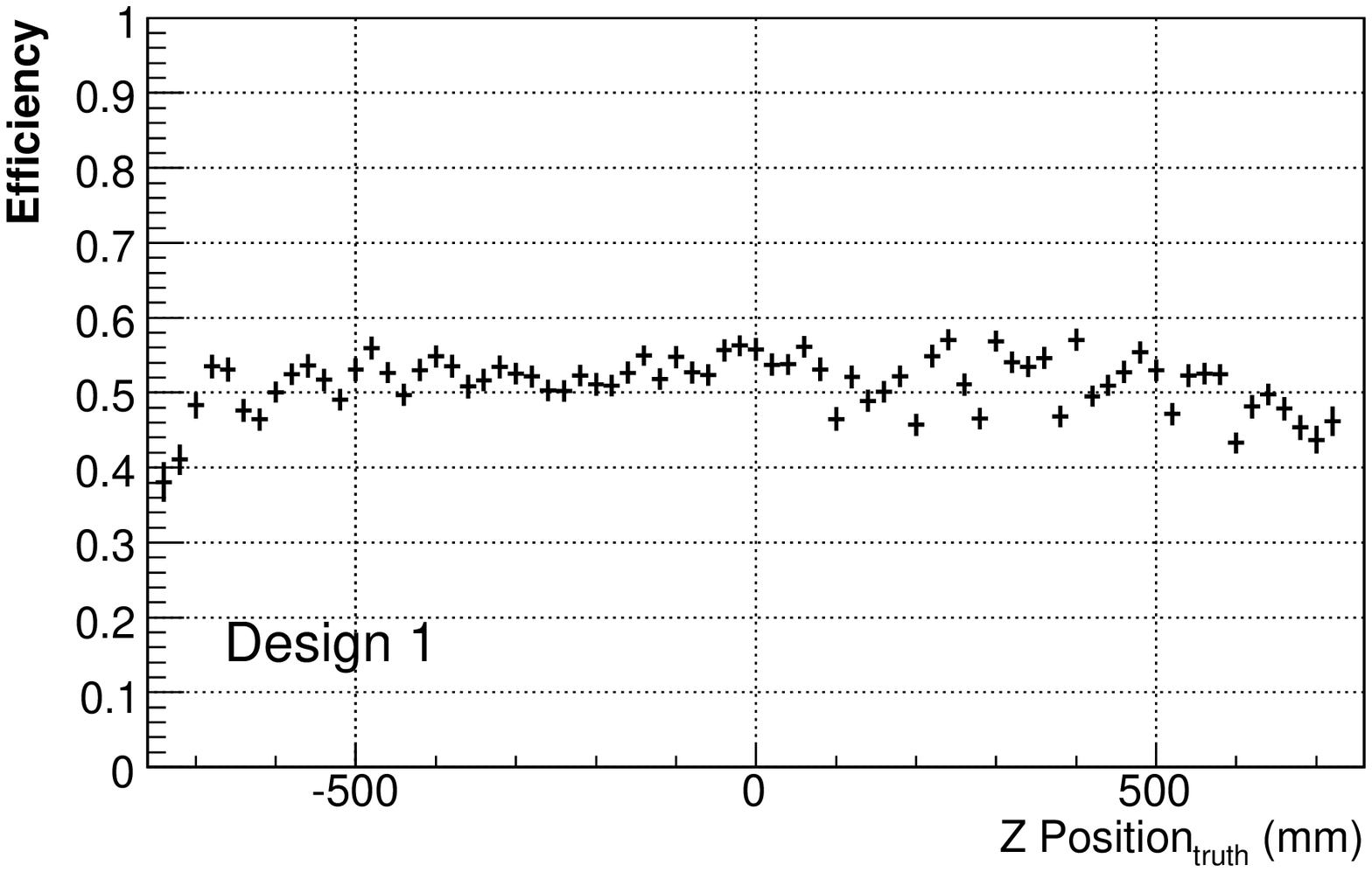}
\hfill
\includegraphics[width=\columnwidth]{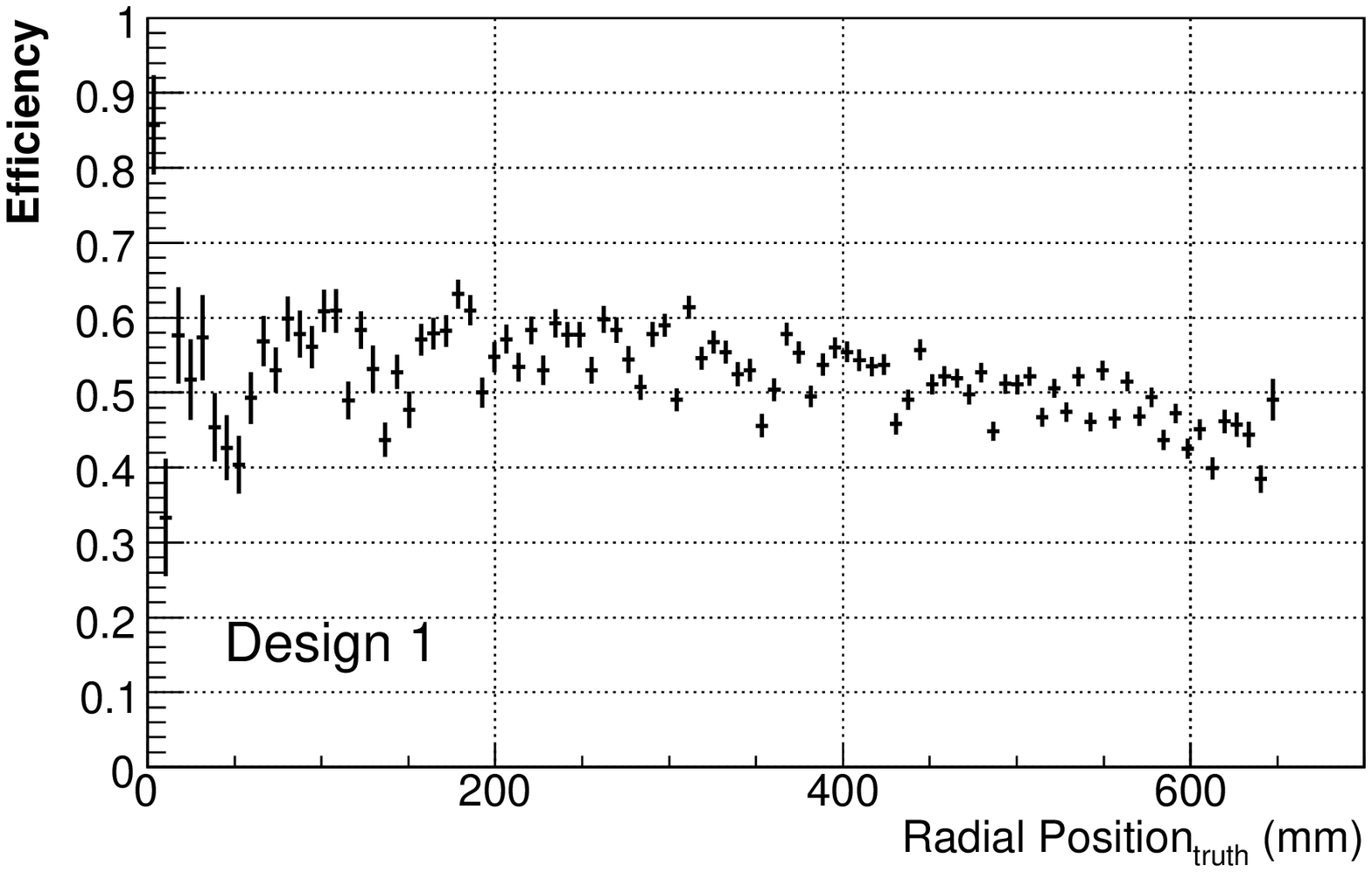}

\includegraphics[width=\columnwidth]{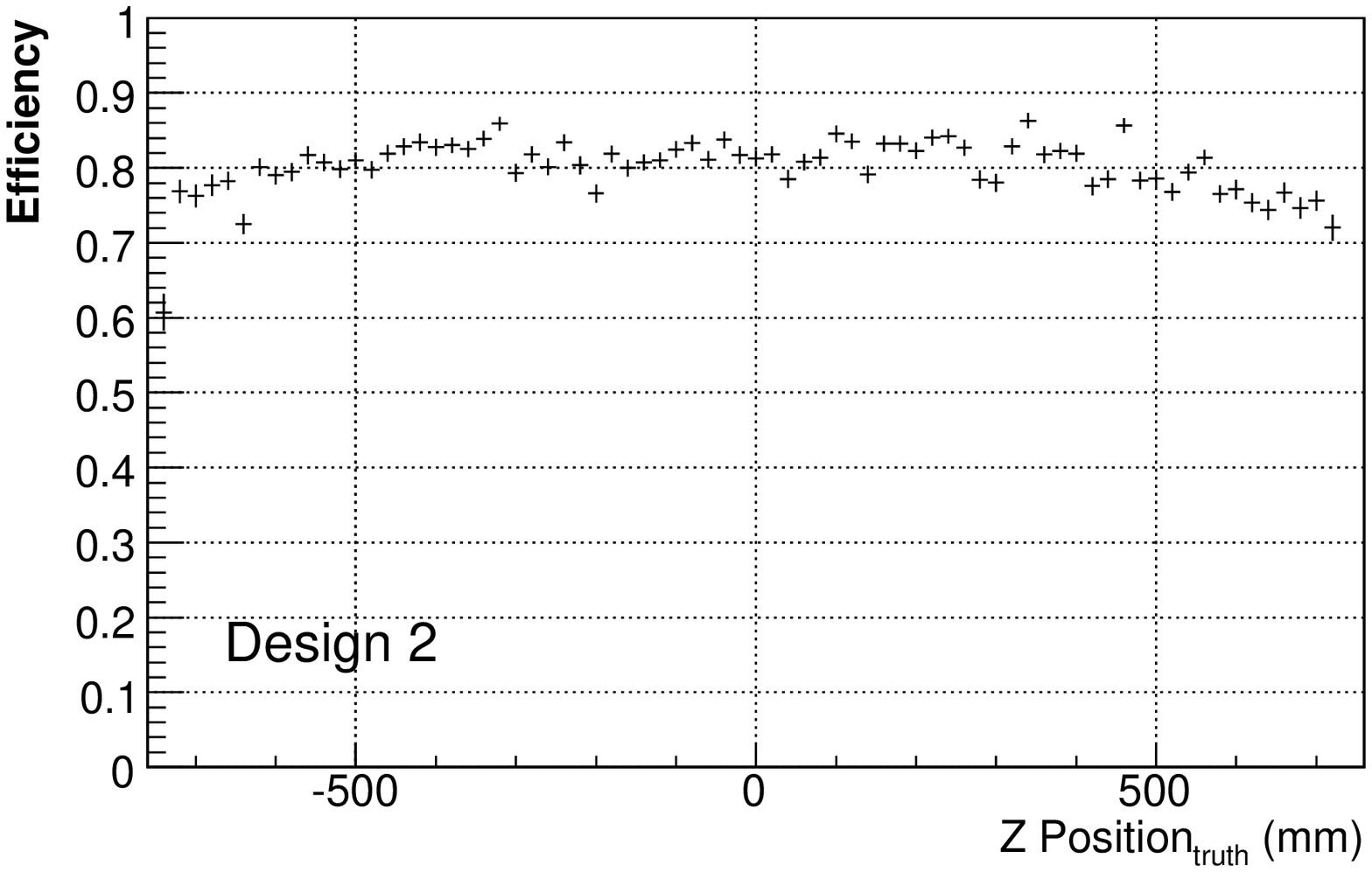}
\hfill
\includegraphics[width=\columnwidth]{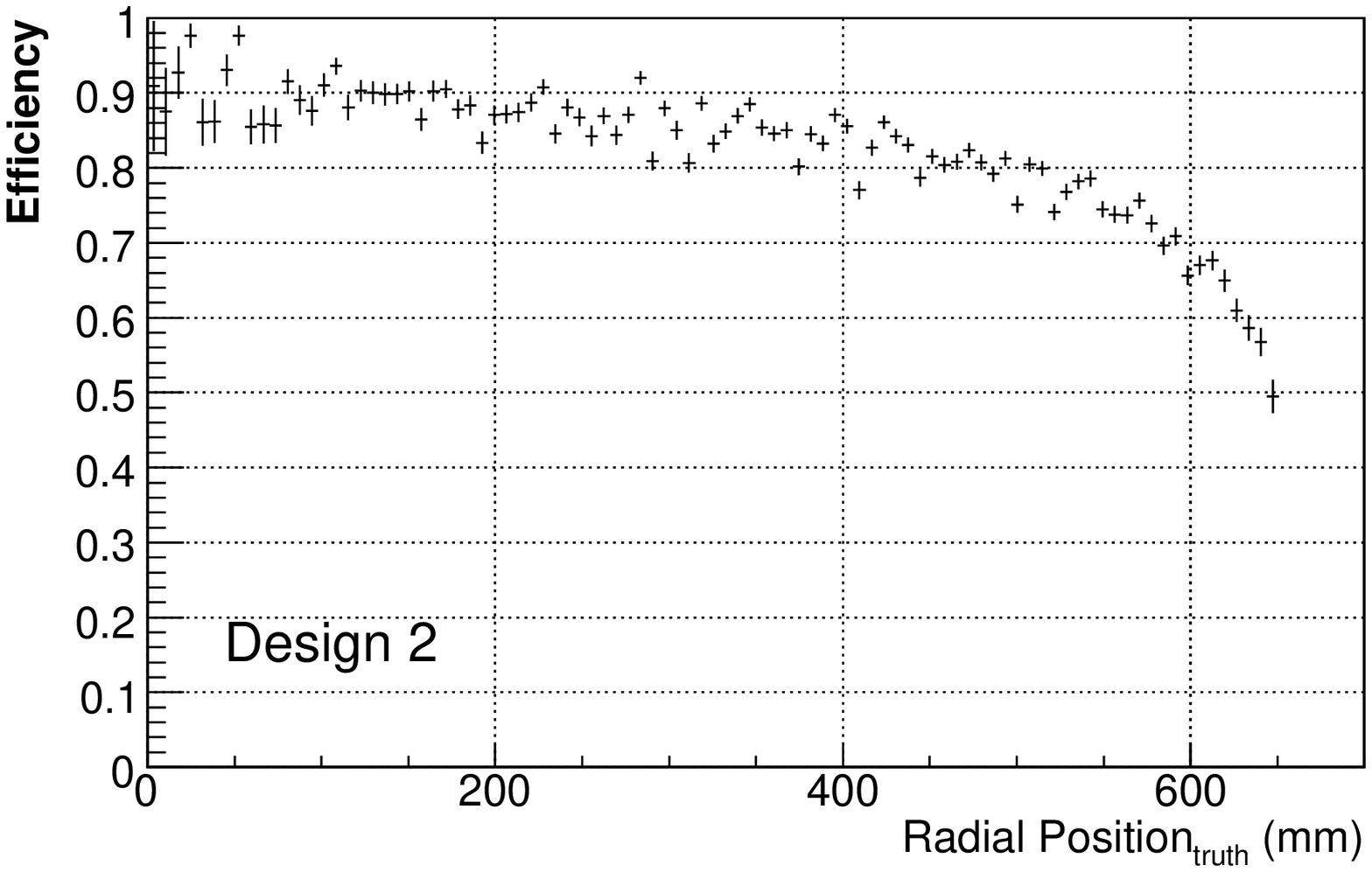}

\includegraphics[width=\columnwidth]{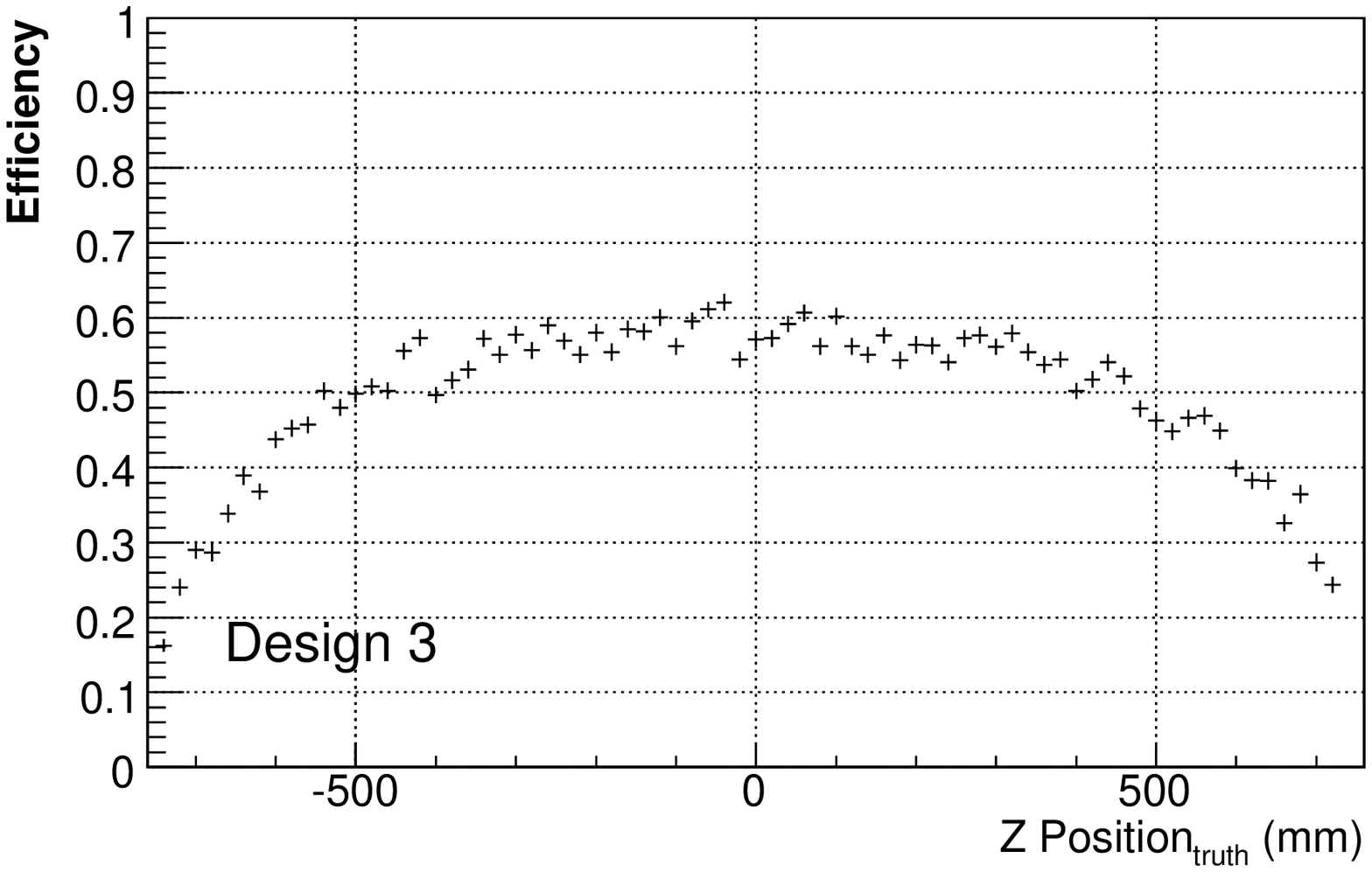}
\hfill
\includegraphics[width=\columnwidth]{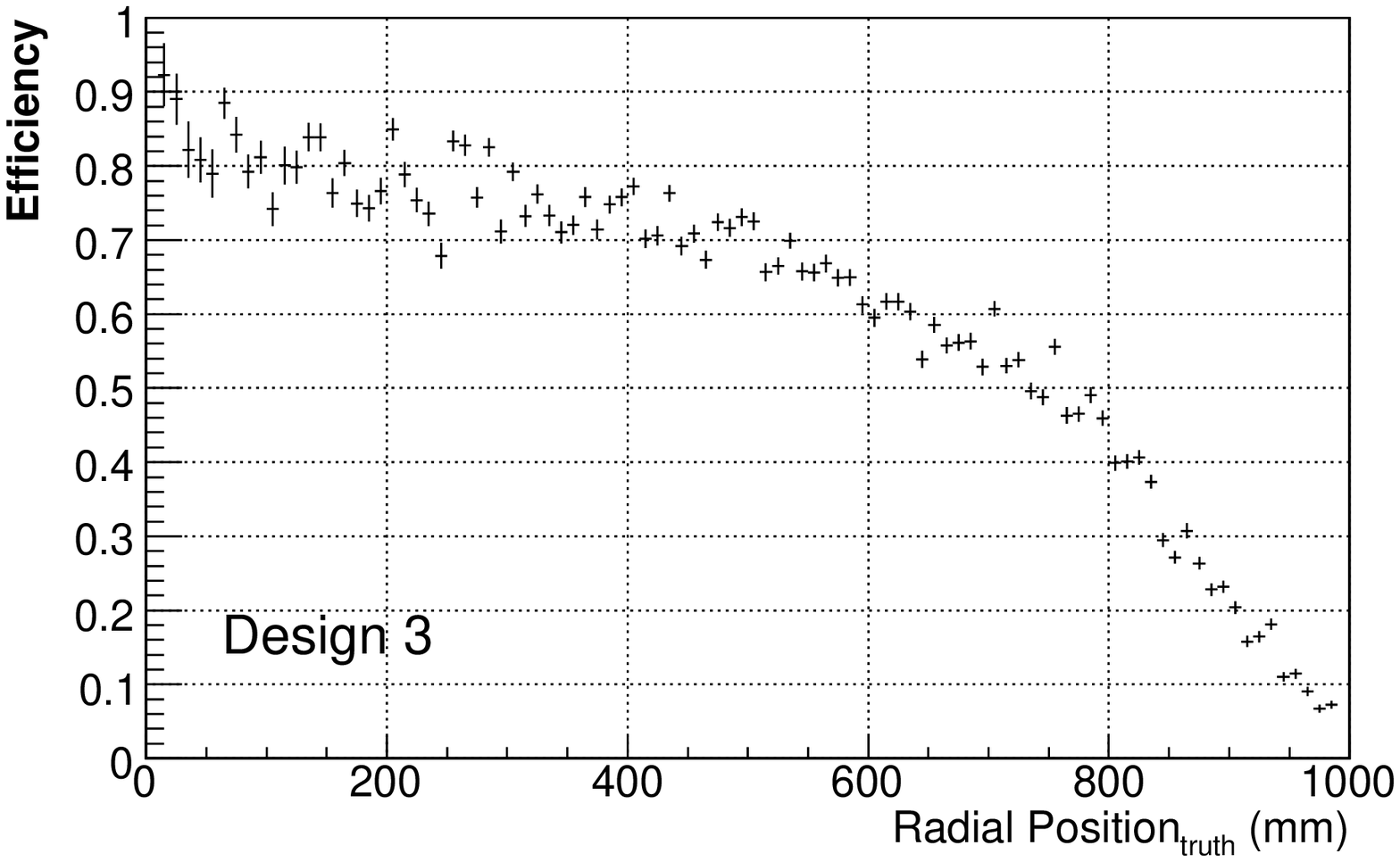}

\includegraphics[width=\columnwidth]{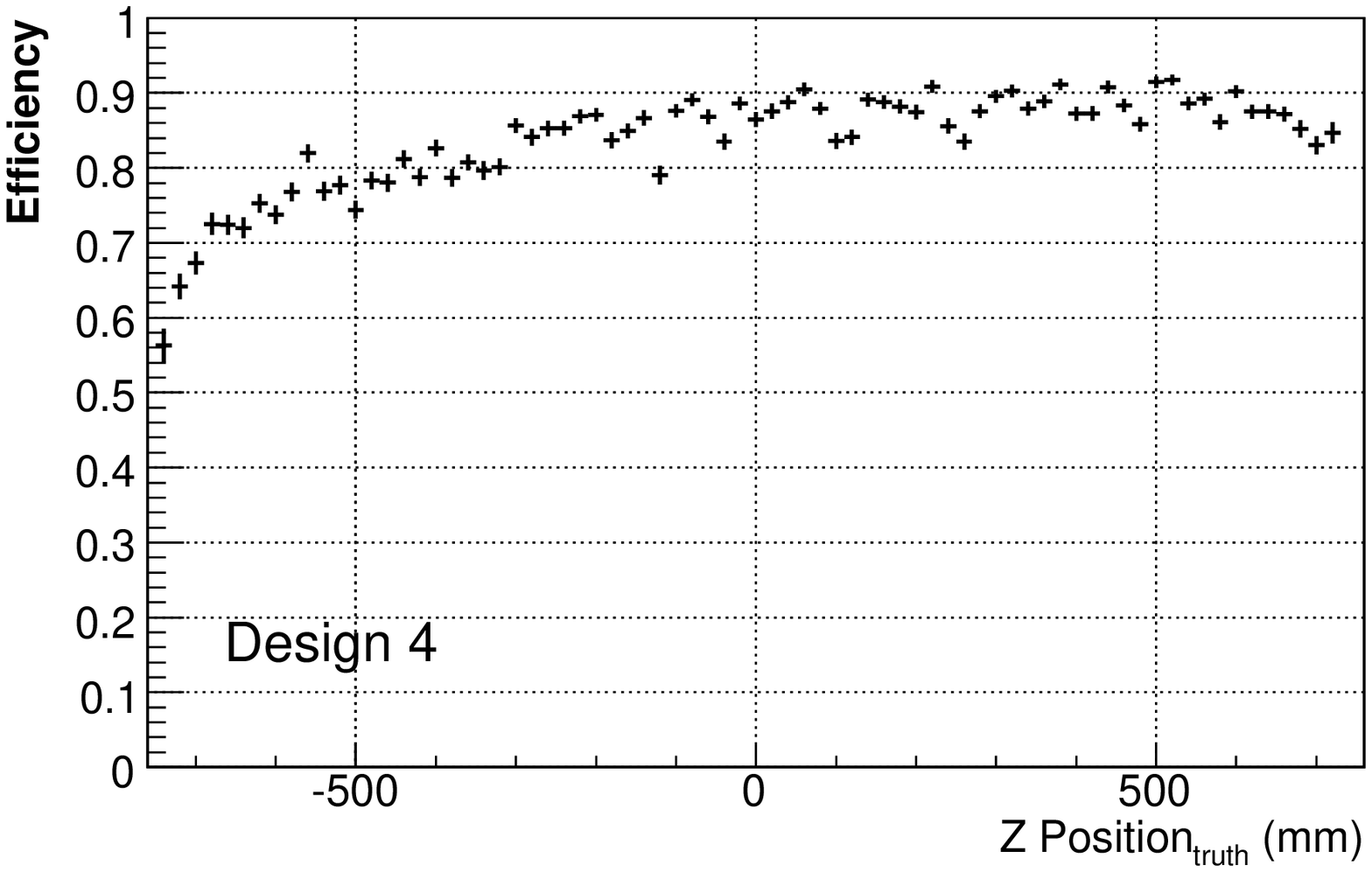}
\hfill
\includegraphics[width=\columnwidth]{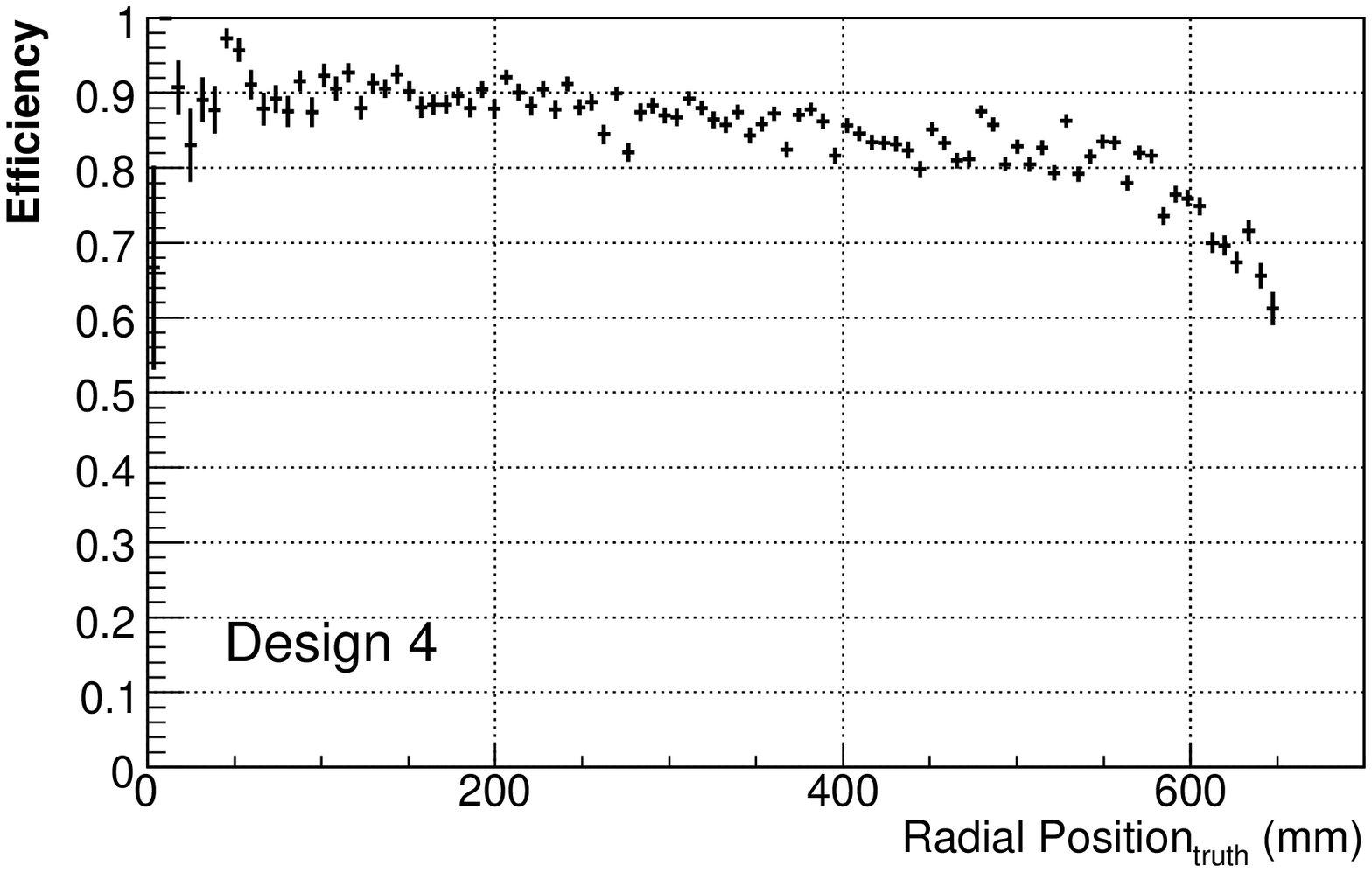}

\caption{Neutron identification efficiency as a function of the position
within the target volume for each detector design, as labeled.  
The truth information from the 
montecarlo was used to identify the Z positions (left) and the radial
positions (right) for each neutron capture event.}
\label{F-nCapDist}
\end{figure*}

To better understand the specific characteristics of each design, we 
investigated the dependence of the neutron capture identification efficiency
on position within the target volume.  The truth information from the
montecarlo simulation was used to identify the vertex position of the 
neutron capture.  The events were binned in the vertical (z-position)
and radial distances from the center of the target volume.  Using the same
threshold described above for the definition of an accepted neutron 
capture event, the efficiencies, as a function of both z-position and 
radius, are plotted for all detector designs in Fig.~\ref{F-nCapDist}.

As expected, one can quite clearly see that, despite the low overall
efficiency, Design~1 shows the most uniform response.  In a similar 
fashion, Design~3, with the lack of any gamma-catcher, shows significant
degradation in the efficiency as it nears the boundaries of the
target volume.  Perhaps more interesting is to compare the radial
performance of Design~2 and Design~3.  Their efficiencies are relatively
similar out to the nominal radius of just over 600~mm.  However, in 
the additional target volume of Design~3 (radius $>$ 650~mm) the efficiency 
drops off quit radically.  Perhaps the most surprising observation
is the relatively good performance of Design~4.  Its z-position dependence
is not as good as Designs 1 or 2, but for most of the range, its 
efficiency for accepting neutron capture events is significantly higher---
10\% higher than Design~2 and 80\% higher than Design~1.  It is only in
the bottom 25--30~cm that the efficiency of Design~4 drops off to values
equal to those of Design~2.  Radially, Design~4 appears to maintain better
uniformity than all other designs with the exception of Design~1, only
showing a declining efficiency for the last 10~cm.  

Of course, most of the volume of the target cylinder is at the larger 
radii where the efficiencies are dropping off.  In fact, the efficiencies 
shown in Table~\ref{T-eff} are under-weighted at large radii due to the 
loss of 2--3\% of the generated neutrons which have leaked out of the 
fiducial volume.  In a real detector, these events would be compensated by
an almost equal number of events, occurring outside of the fiducial
volume, for which neutrons would be captured within the target region.
To better evaluate the total effective volume of each of these
designs, we assumed a uniform distribution of neutron captures and
integrated the differential efficiencies over the total volume
of each design's target region with the following results: 
Design~1~=~0.93~m$^3$; Design~2~=~1.42~m$^3$; Design~3~=~2.15~m$^3$; 
Design~4~=~1.51~m$^3$.  Recall that the nominal target volume was 2~m$^3$
and that Design~3 actually had a volume of 4.72~m$^3$.

\section{Positron Energy Response}
The positrons from a reactor induced inverse beta-decay event will deposit
between 1--6~MeV of energy into the scintillator with a maximum likelihood
value of $\sim$2.5~MeV.  We performed some simple studies of the
linearity of the energy response to positrons with kinetic energies between
0--5~MeV and did not see any significant energy dependence.  We therefore
concentrated on two specific positron samples---one with kinetic energy of
1.5~MeV corresponding to the peak energy deposition and one with
kinetic energy of 3.5~MeV which would be in the high energy tail---which
we used to study in detail the position dependence of the different 
detector designs.  

\begin{figure}[htb]
\includegraphics[width=\columnwidth]{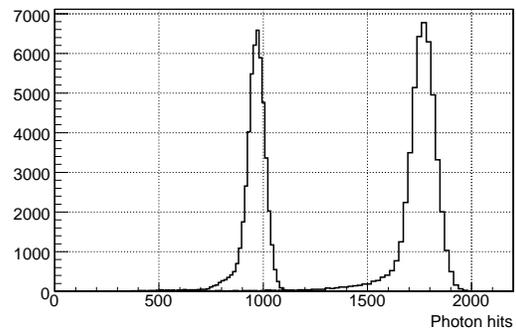}
\caption{Total detected energy deposition for positrons generated within the 
target volume of detector Design~1.  The photon hits represent the simulated
number of photoelectrons at the PMTs.  The positrons were generated with
kinetic energies of 1.5 and 3.5~MeV.}
\label{F-posDepD1}
\end{figure}

Each positron sample was generated uniformly within the target volumes of each
detector design.  The simulation libraries described above were used to 
montecarlo the ionization losses and the eventual positron annihilation.
The ensuing 
optical photons from the scintillator were tracked to the photocathodes of
the PMTs.  Within the simulation, the PMTs used a radially dependent 
quantum efficiency which is consistent with experimental testing.  This 
yielded a photon hit-count per PMT that is roughly equivalent to a number
of photoelectrons.  This photon hit-count has been used for all comparisons
of detector response---avoiding the complications of varying electronics
and data acquisition systems.  All PMT hits recorded within 100ns were summed
together to create an effective total charge deposition. An example of the 
total energy deposition from these two samples in Design~1 is shown in 
Fig.~\ref{F-posDepD1}.

\begin{figure*}[htbp]

\includegraphics[width=\columnwidth]{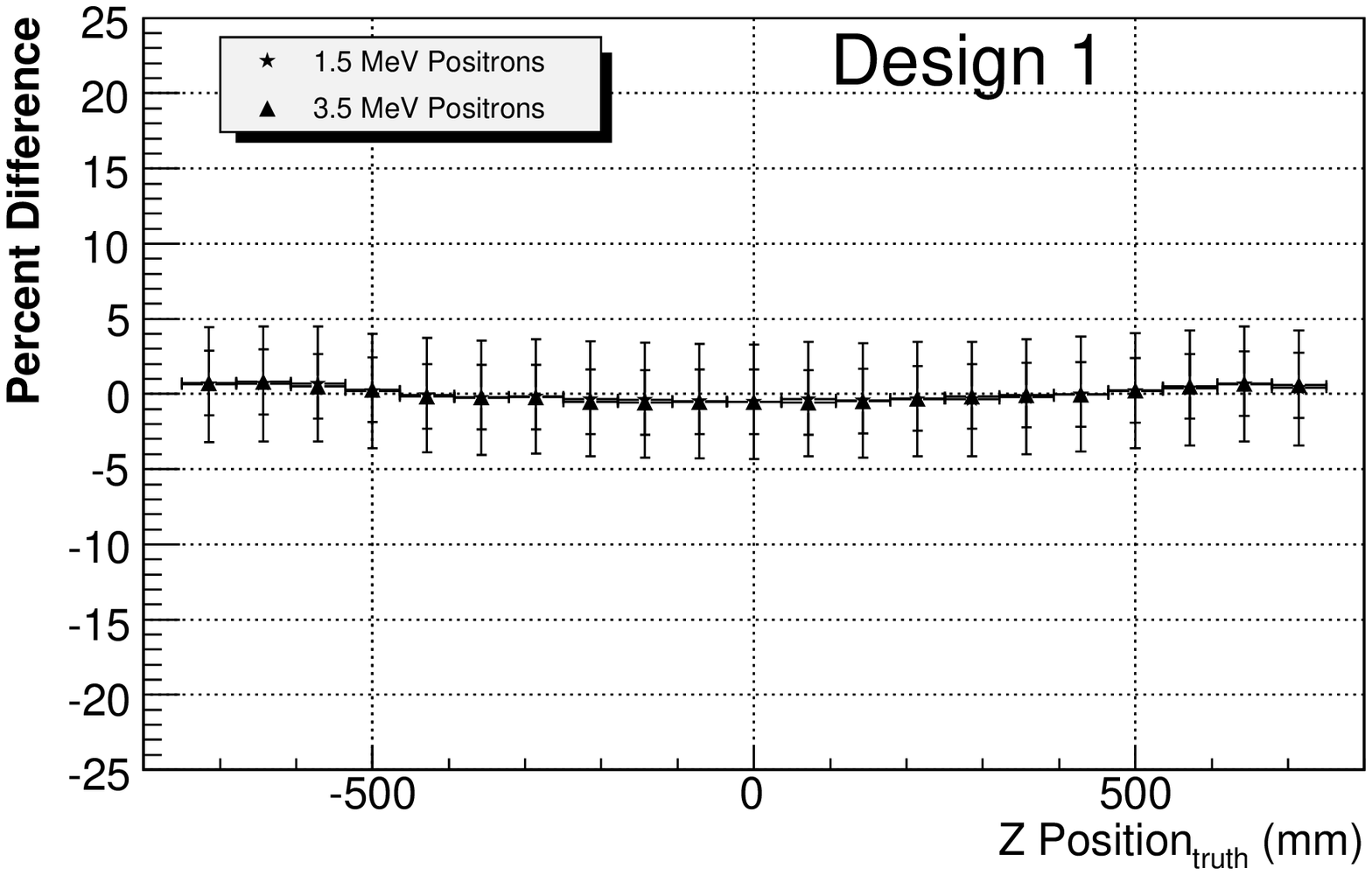}
\hfill
\includegraphics[width=\columnwidth]{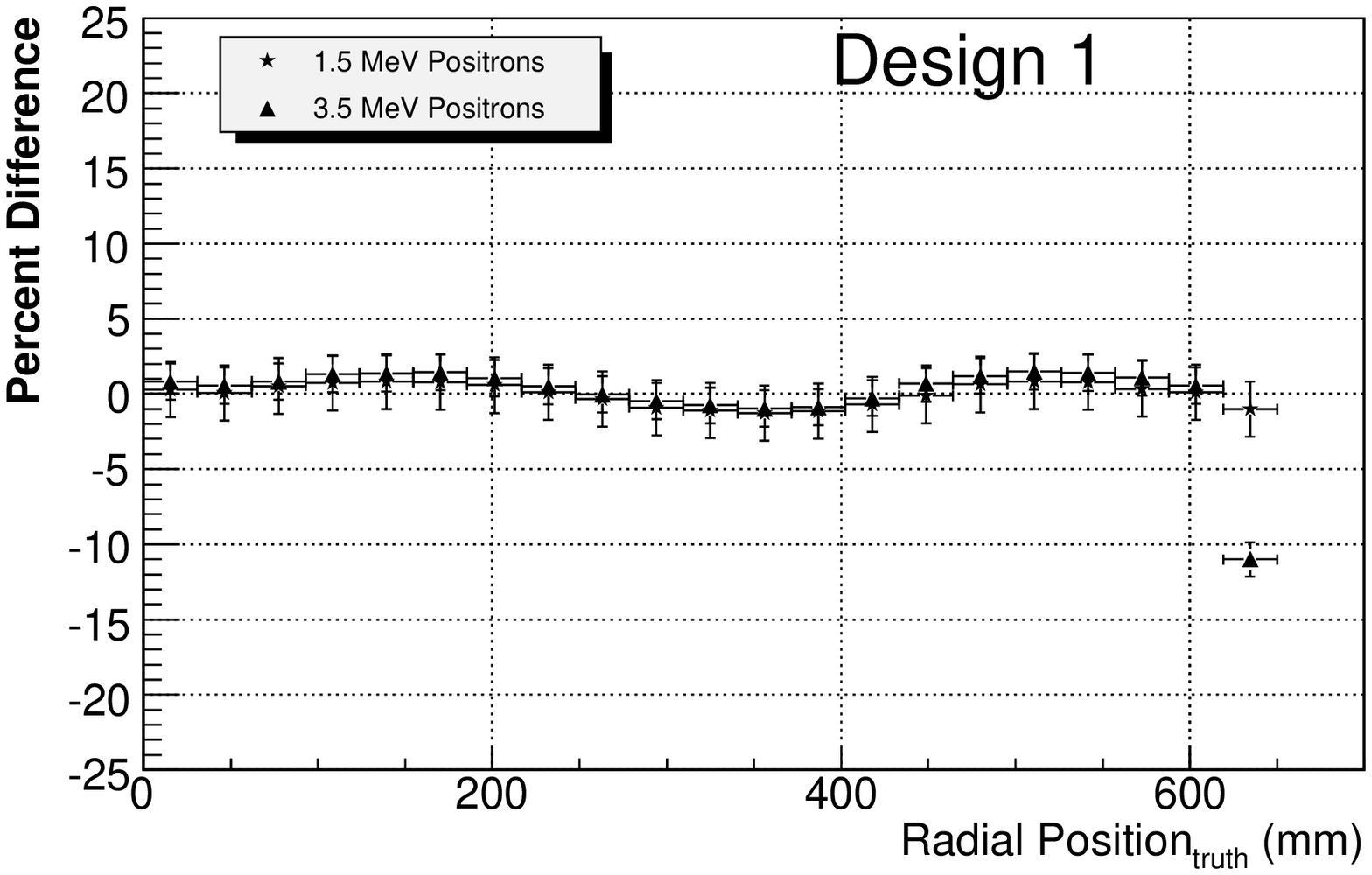}

\includegraphics[width=\columnwidth]{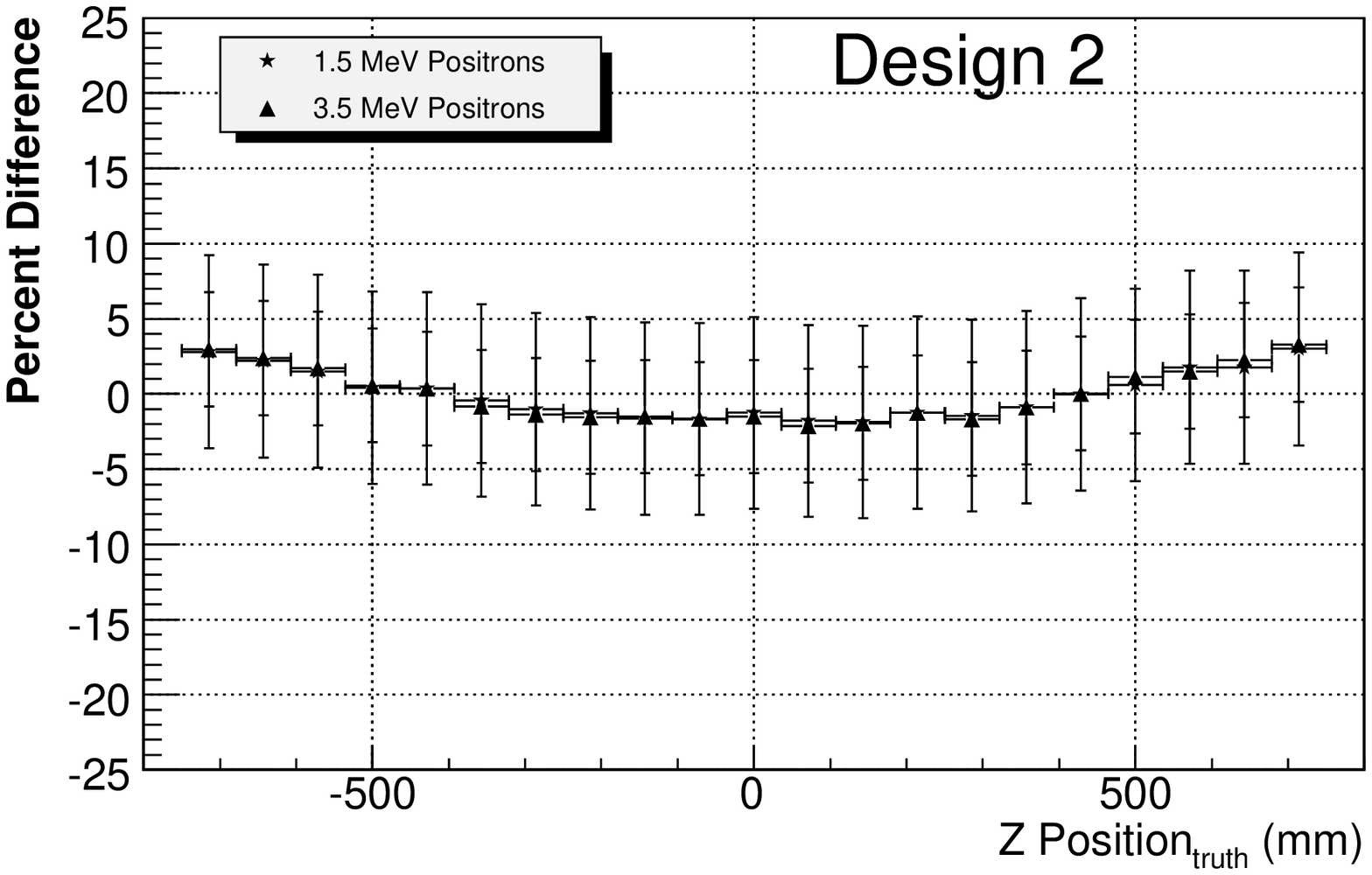}
\hfill
\includegraphics[width=\columnwidth]{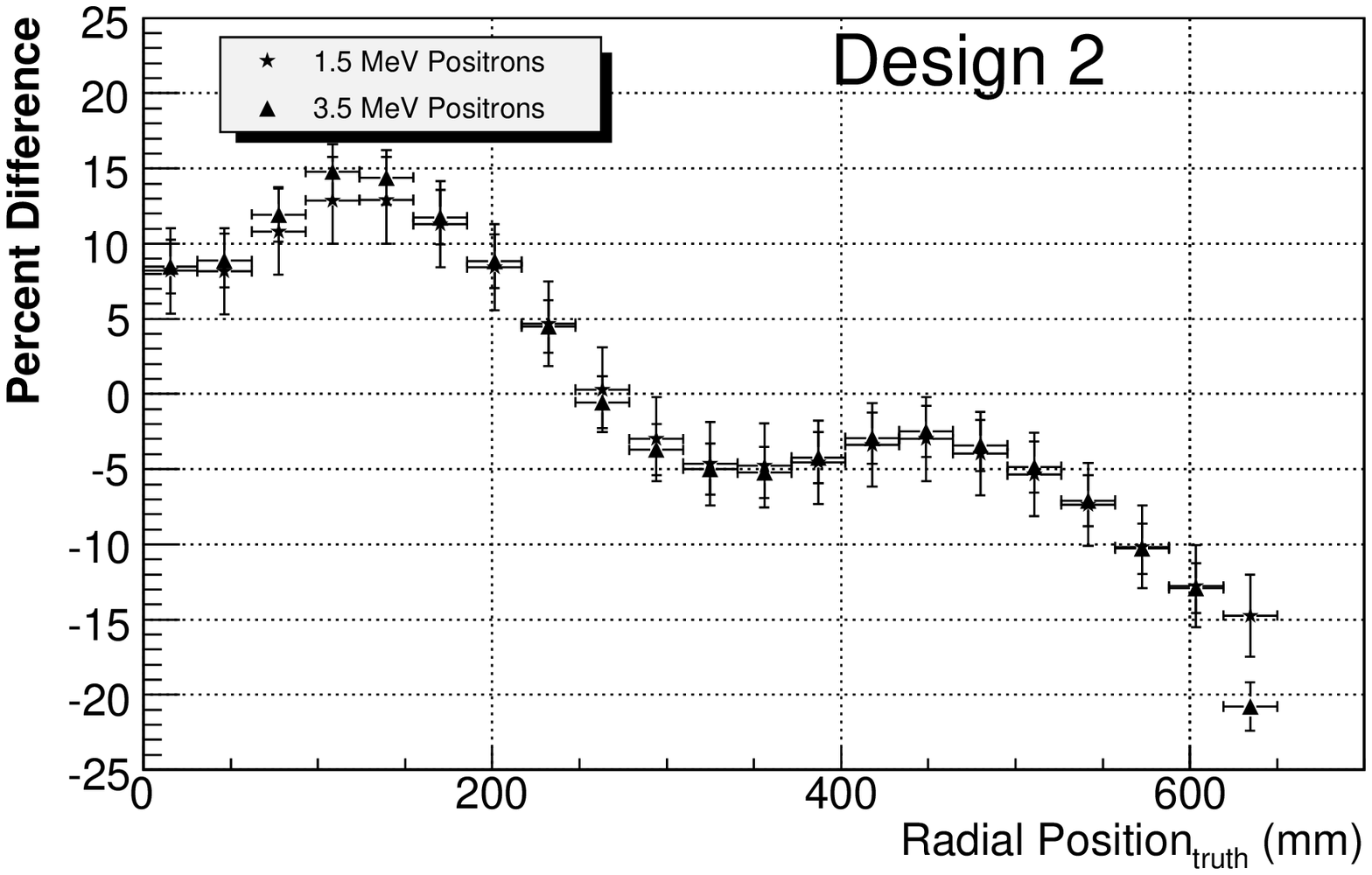}

\includegraphics[width=\columnwidth]{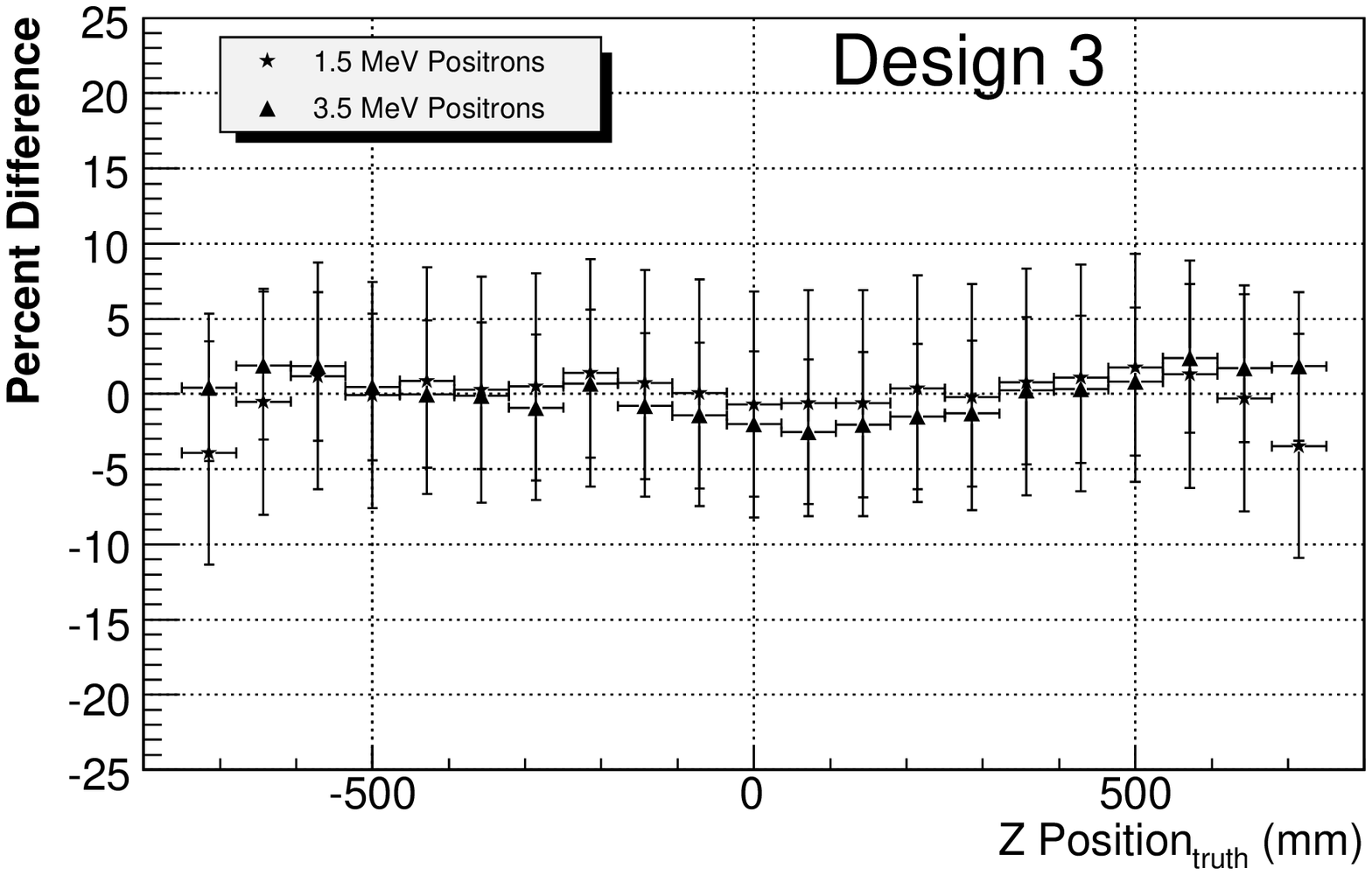}
\hfill
\includegraphics[width=\columnwidth]{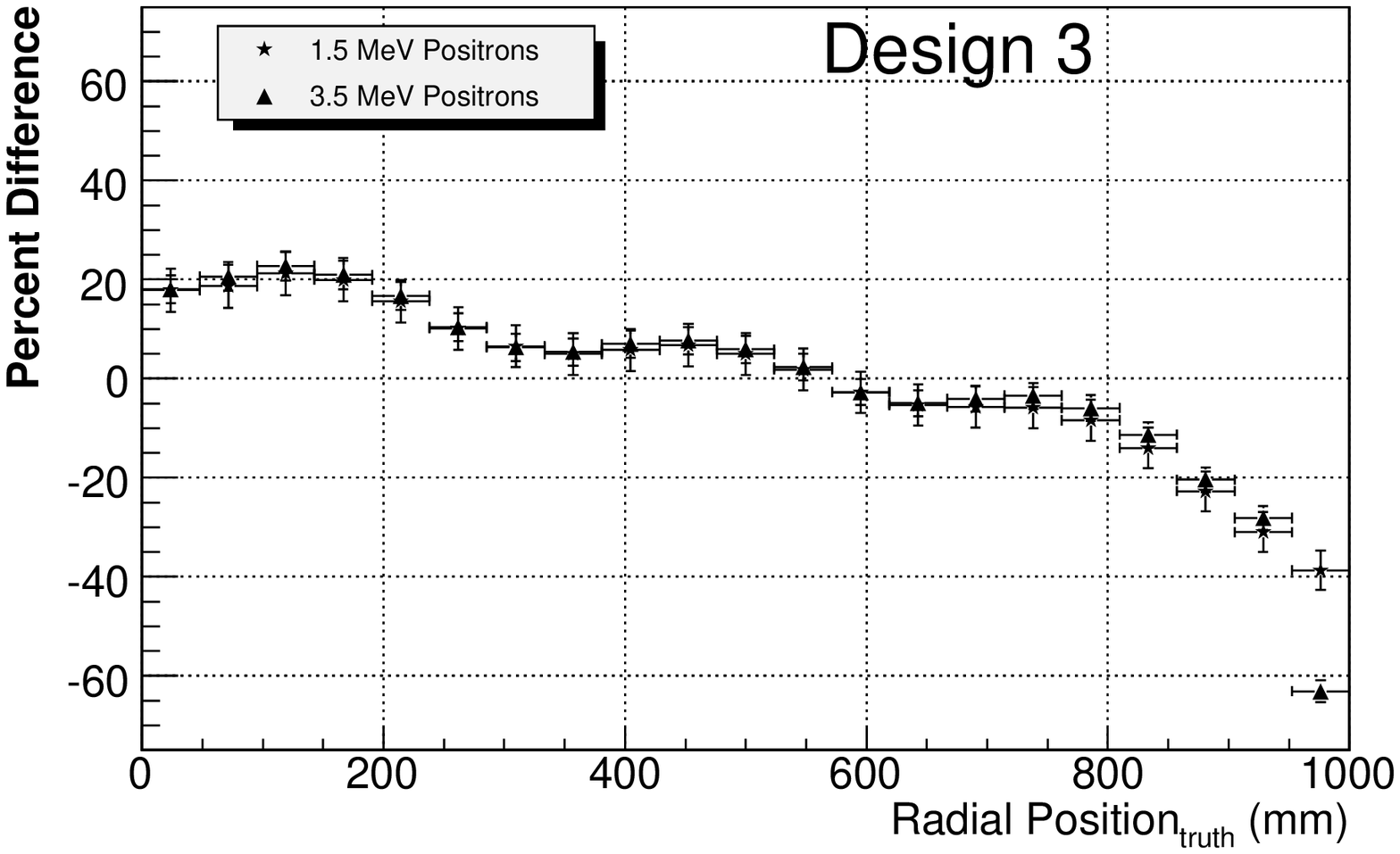}

\includegraphics[width=\columnwidth]{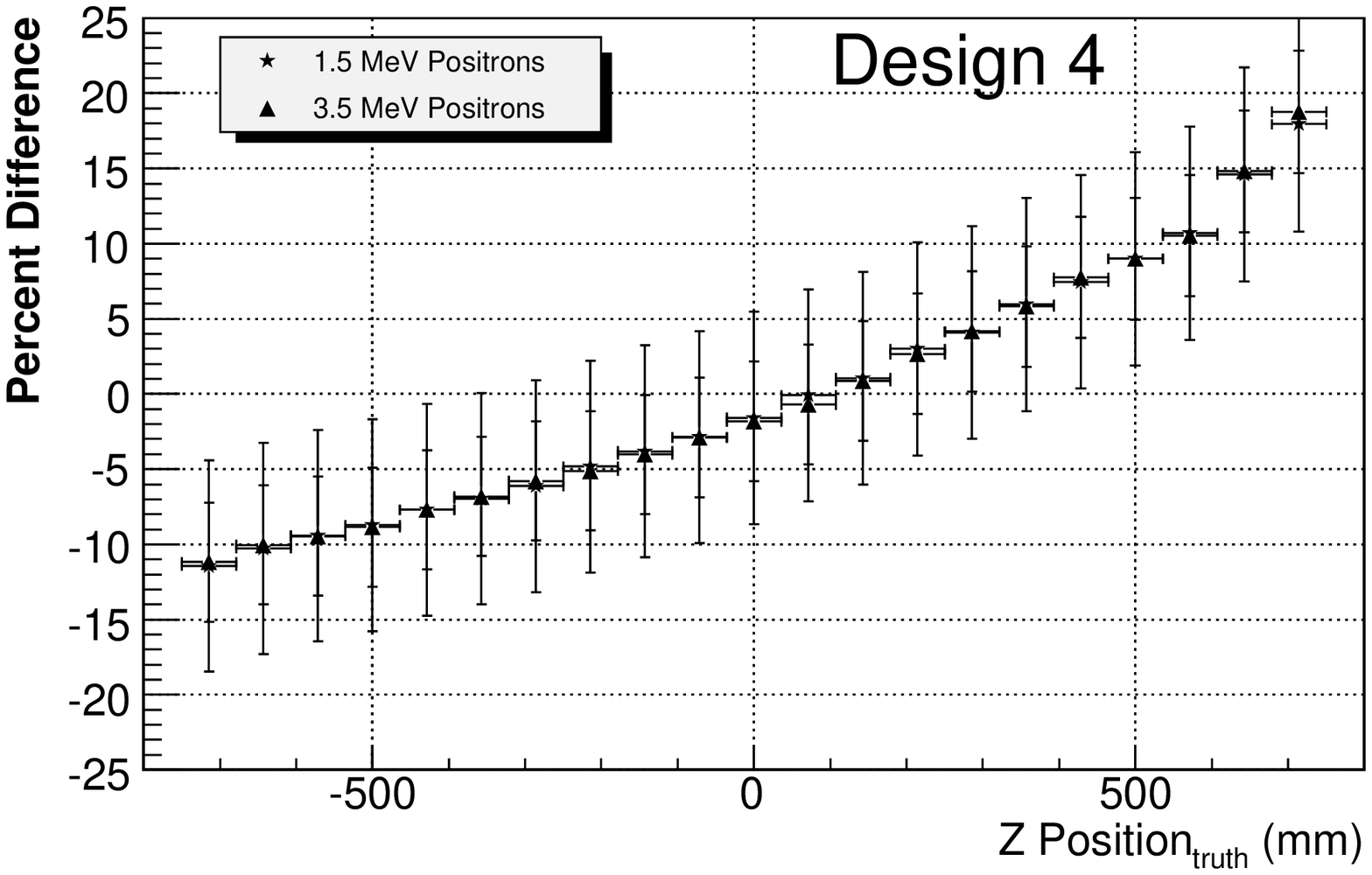}
\hfill
\includegraphics[width=\columnwidth]{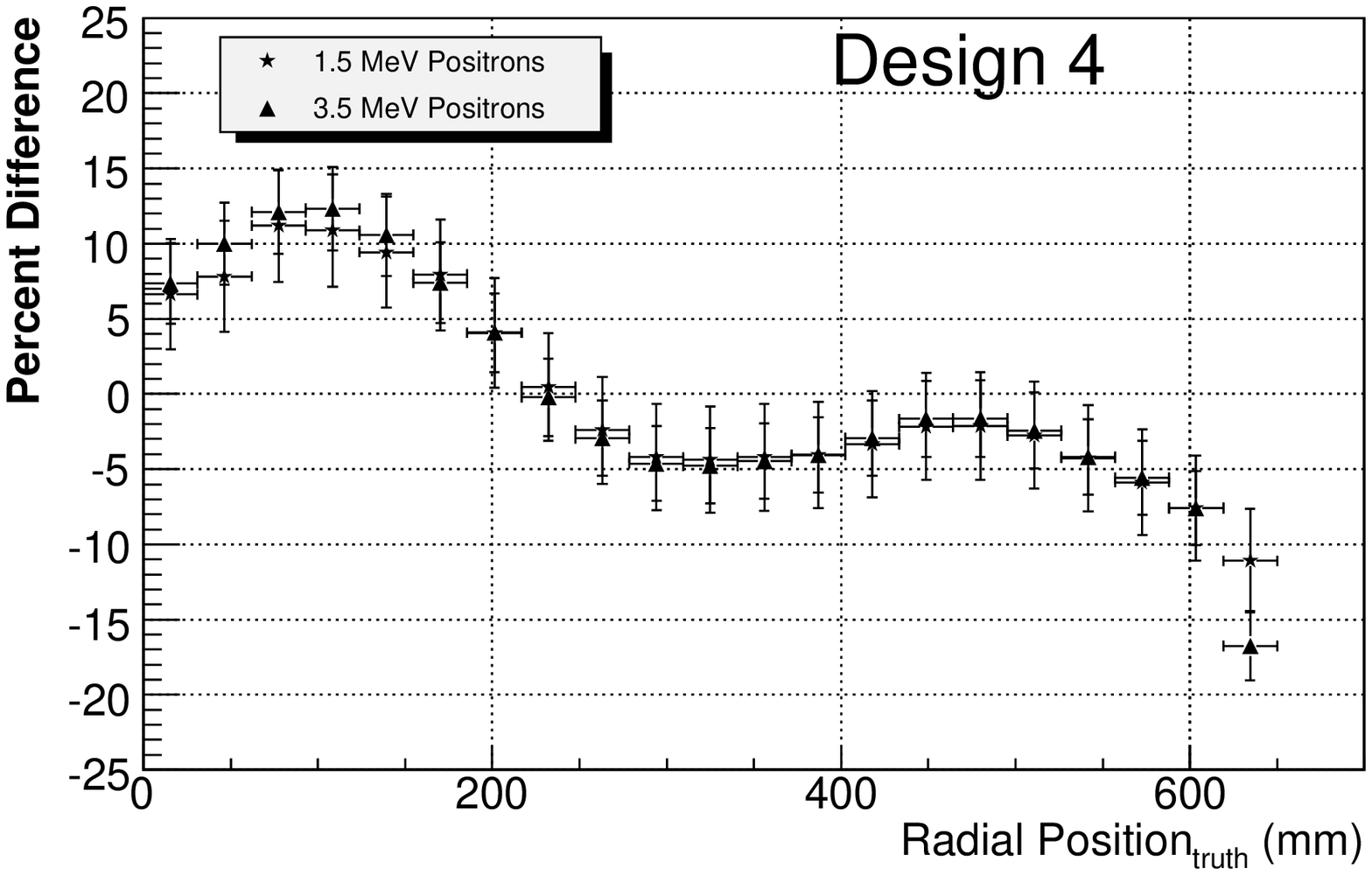}

\caption{Relative positron energy response as a function of position
within the target volume for each detector design, as labeled.  
The truth information from the 
montecarlo was used to identify the Z positions (left) and the radial
positions (right) for each positron event.  The responses of positrons with 
kinetic energy of 1.5 and 3.5~MeV are shown simultaneously as a percent 
difference from the mean response of the detector.  Errors represent the
fitted Gaussian sigma of the data in each bin.}
\label{F-posDist}
\end{figure*}

The truth information from the
montecarlo simulation was used to identify the vertex position of the 
positron event.  The events were binned in the vertical (z-position)
or radial distances from the center of the target volume.  The recorded 
photon hit-counts were fitted to a Gaussian for each bin in either 
radial distance or z-position.  These data, for all detector designs,
are shown in Fig.~\ref{F-posDist} as a percent difference from the 
mean response of the detector.  The errors shown represent the Gaussian
sigma of the fit.

The first thing to notice when looking at these distributions is that
the responses to the 1.5~MeV and 3.5~MeV positrons is quite nearly
identical.  Also, the errors are consistent with the expected energy resolution
from the liquid scintillator.  As with the neutron capture efficiencies,
Design~1 shows the most uniform response of all detector designs.  It
was a bit surprising to see the large variations in the radial responses
for Designs~2--4.  After some further investigation, it was understood that
this structure is consistent with the ring-like implementation of the
PMTs at the top and bottom of the detectors.  The additional PMTs on
the barrel of Design~1 are clearly serving to reduce these variations.
Also, as was expected, the detector designs which incorporate two-sided
readouts (Designs~2--3) show more uniform response relative to z-position
than is seen in Design~4.  However, given that the radial responses of
these detector designs show such large variations, it is not clear that
the z-position uniformity provides a significant gain.  In fact, it 
appears that Design~4, with a slightly larger surface area of PMTs at
the single end, provides a more uniform response with respect
to radial position than Design~2--- $\pm$12\% compared to $\pm$15\%.
In addition, this appears well matched to the z-dependent response of
Design~4 which is relatively similar.

\subsection*{Simple Energy Calibration}
An attempt was made to see if the variations in detector response could
be reduced by implementing a simple energy calibration as a function of 
vertex position.  Clearly Design~1, while providing sufficient PMTs
to reconstruct a vertex position with good accuracy, provides a good
uniformity of response and was therefore not attempted.  On the other
hand, Design~4 demonstrated that readout from only one side was insufficient
to consistently reconstruct a vertex position that would be useful for
correcting the energy response.  

\begin{figure}[htb]
\includegraphics[width=0.75\columnwidth]{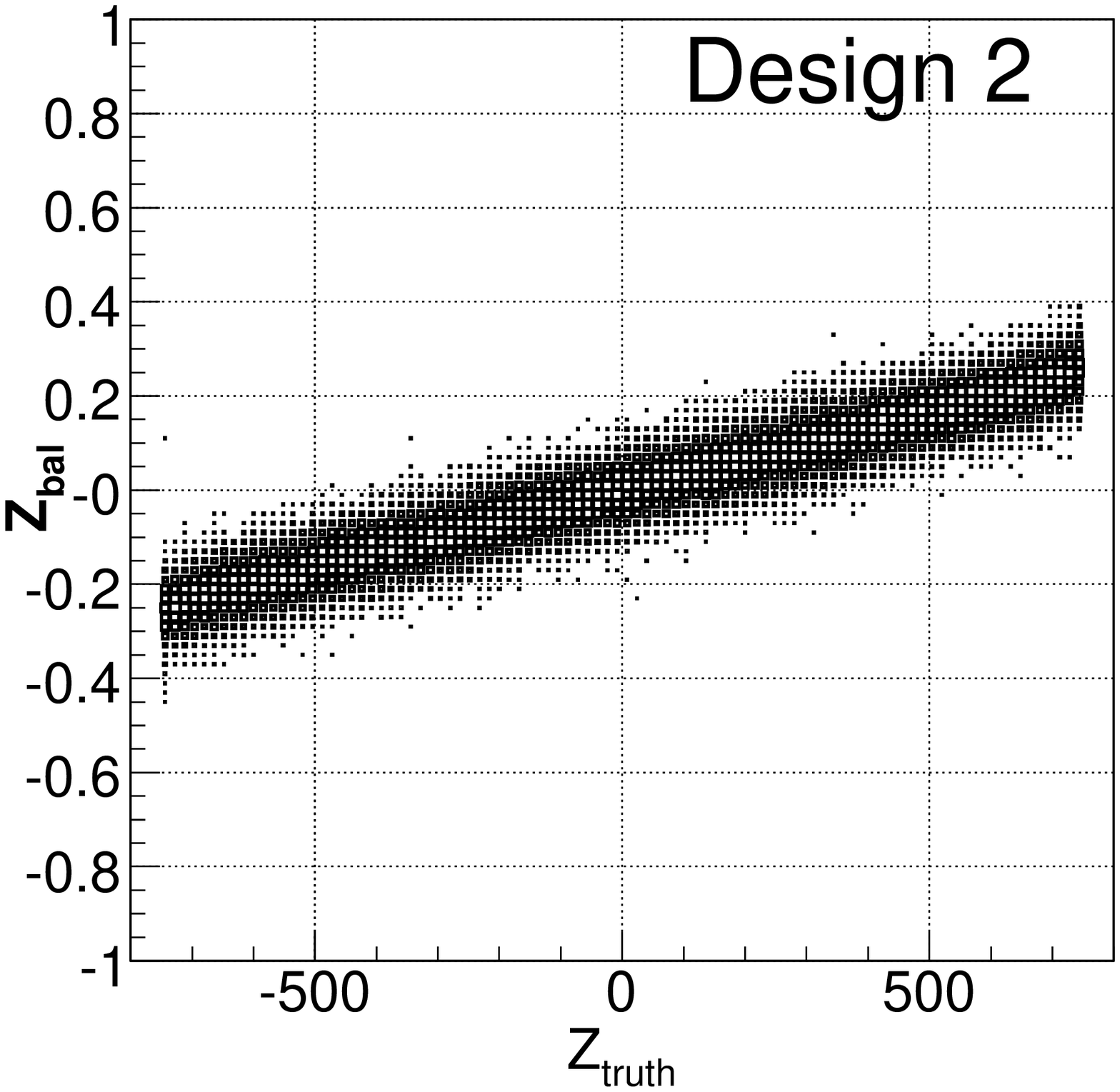}
\includegraphics[width=0.75\columnwidth]{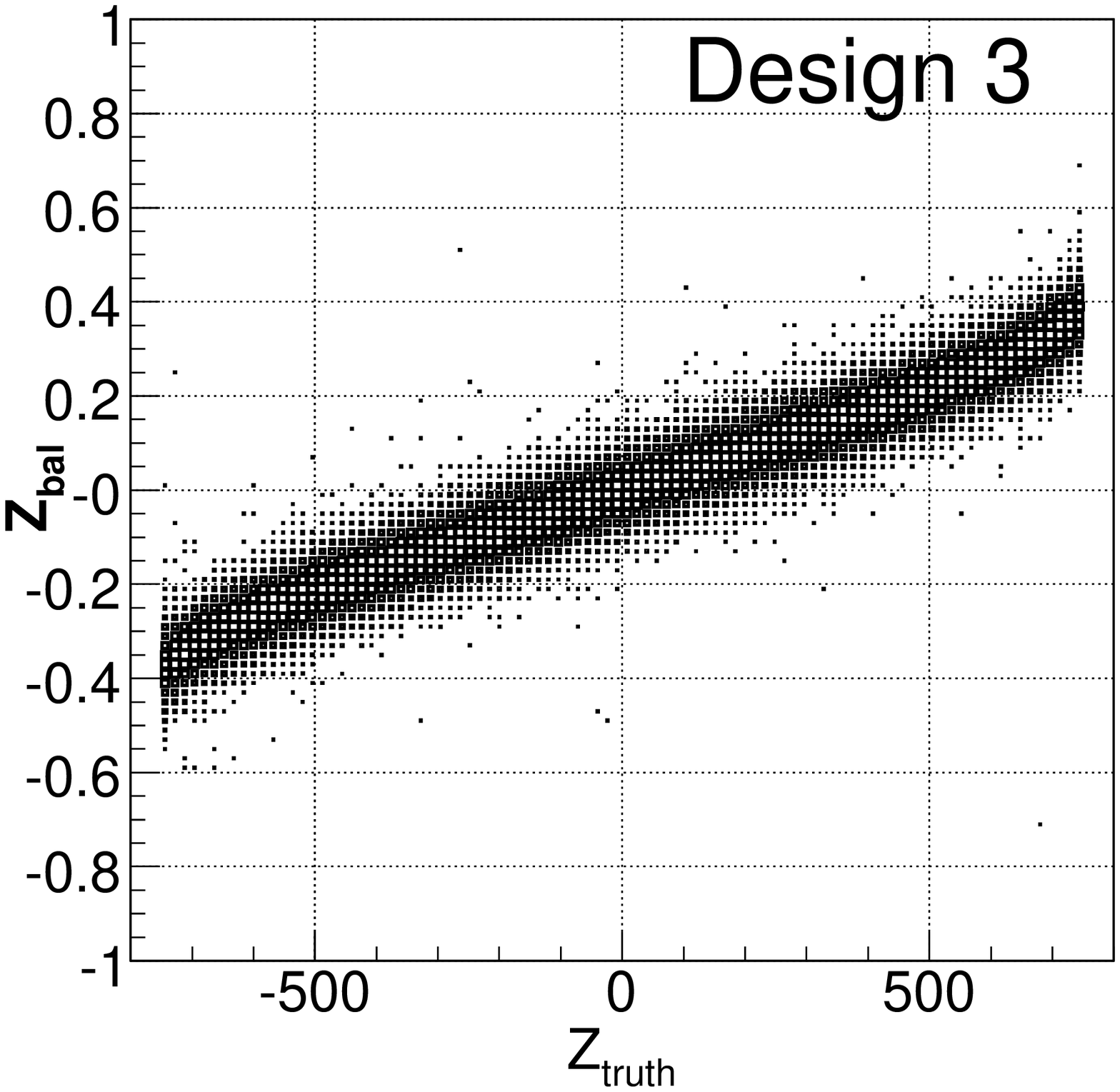}
\caption{Correlation between the measurable z-balance, as defined in
Eq.~\ref{E-zBal}, and the true z-position from the montecarlo for 
detector Designs 2 and 3.  While the correlation is relatively linear
for both designs, one can see the non-linearity of Design~3 at the extremes
due to the energy leakage in the absence of a gamma-catcher}
\label{F-zCor}
\end{figure}

\begin{figure}[htb]
\includegraphics[width=0.85\columnwidth]{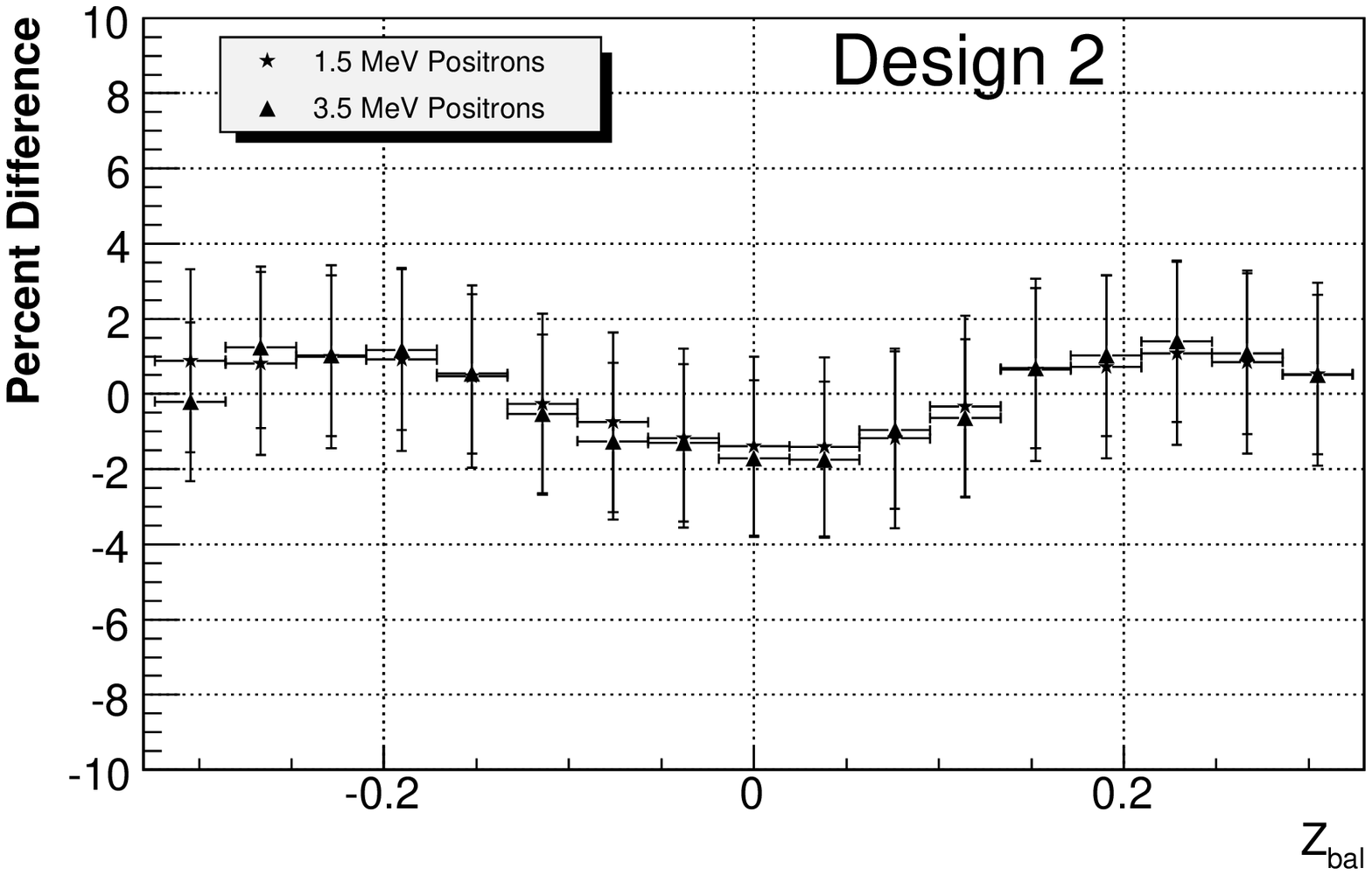}
\includegraphics[width=0.85\columnwidth]{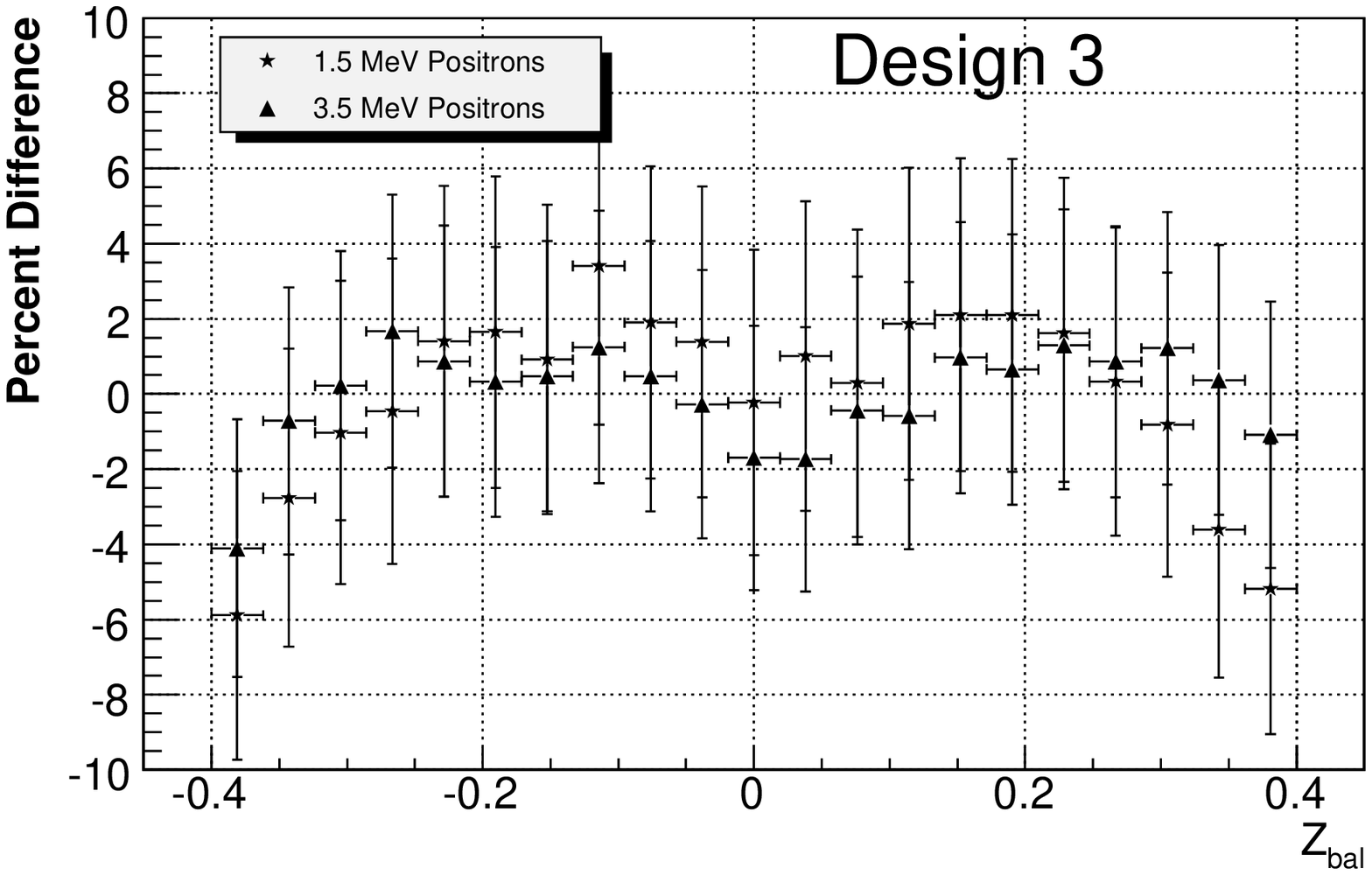}
\caption{Relative positron energy response as a function of the z-balance
as defined in Eq.~\ref{E-zBal} for detector Designs 2 and 3, as labeled. 
While the gross structure of the position dependence is similar to that
shown in Fig.~\ref{F-posDist}, the magnitude of the variation is slightly 
less.}
\label{F-posUncal}
\end{figure}

For Designs~2 and 3, we attempted various methods to reconstruct the
radial position.  However, the reflectivity of the side walls made this
unreasonably difficult.  We did find, however, that a simple charge
balance between the top and bottom PMTs provided a very good correlation
to the true z-position.  We define the measurable quantity z-balance 
($Z_{bal}$) in the following way: 
\begin{equation}
Z_{bal} = \frac{Top - Bottom}{Top + Bottom},
\label{E-zBal}
\end{equation}
where $Top$ and $Bottom$ refer to the total number of photon hits recorded
by the PMTs at the top and bottom of the detector, respectively.  The
correlation of this value to the true z-position for Designs~2 and 3
can be seen in Fig.~\ref{F-zCor} and the relative positron
energy response (similar to Fig.~\ref{F-posDist}) is plotted relative
to this variable in Fig.~\ref{F-posUncal}.
Using the data from Fig.~\ref{F-posUncal}, a multiplicative calibration 
constant for each bin in $Z_{bal}$ was constructed to give the mean
response.  Rather than creating an energy dependent calibration, 
the relative responses of the 1.5 and 3.5~MeV positron samples were averaged
to provide a single calibration constant for each bin.  This procedure
was repeated for both detector designs and the calibration constants
were applied event-by-event.

\begin{figure}[hb]
\includegraphics[width=0.85\columnwidth]{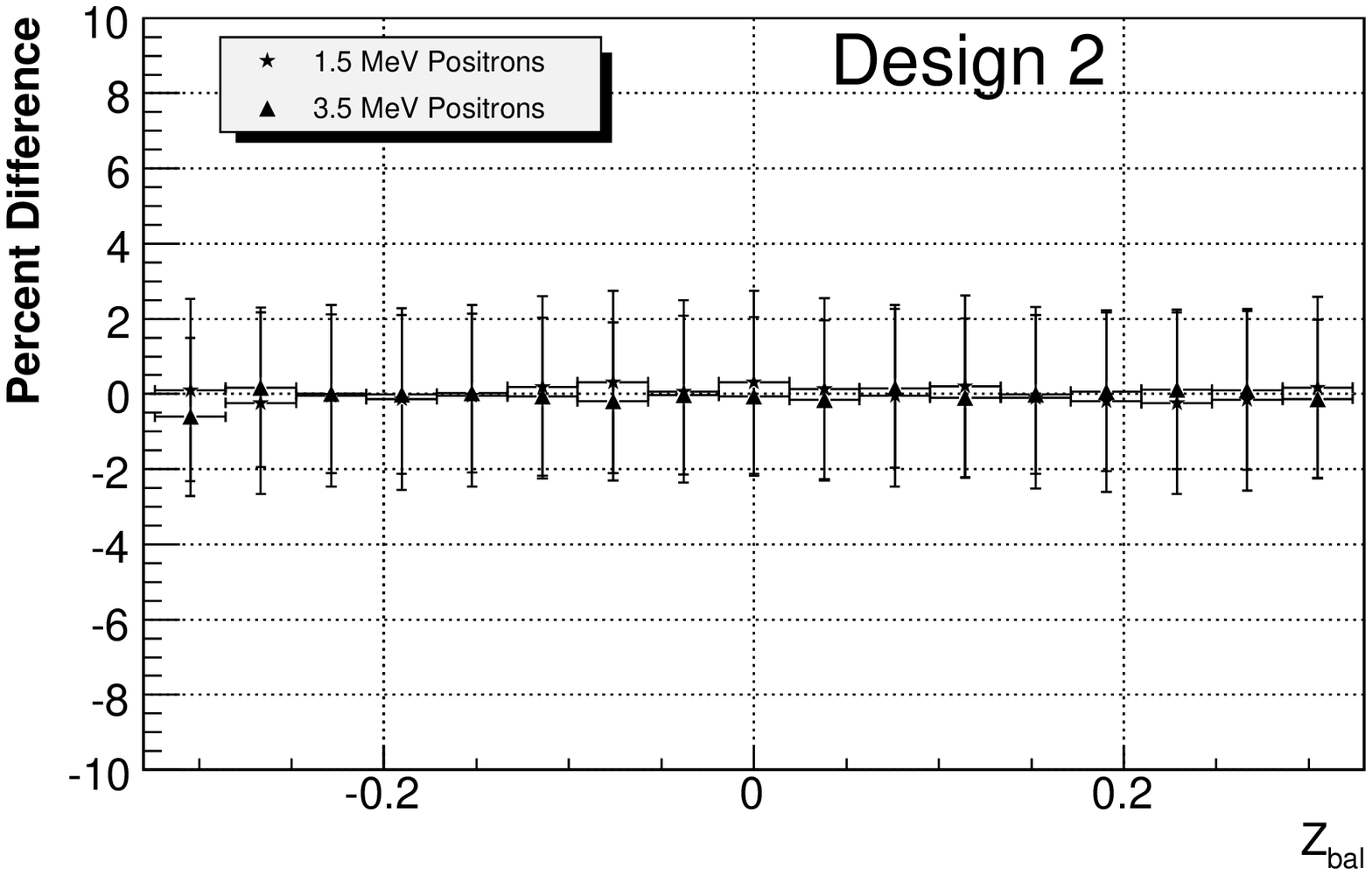}
\includegraphics[width=0.85\columnwidth]{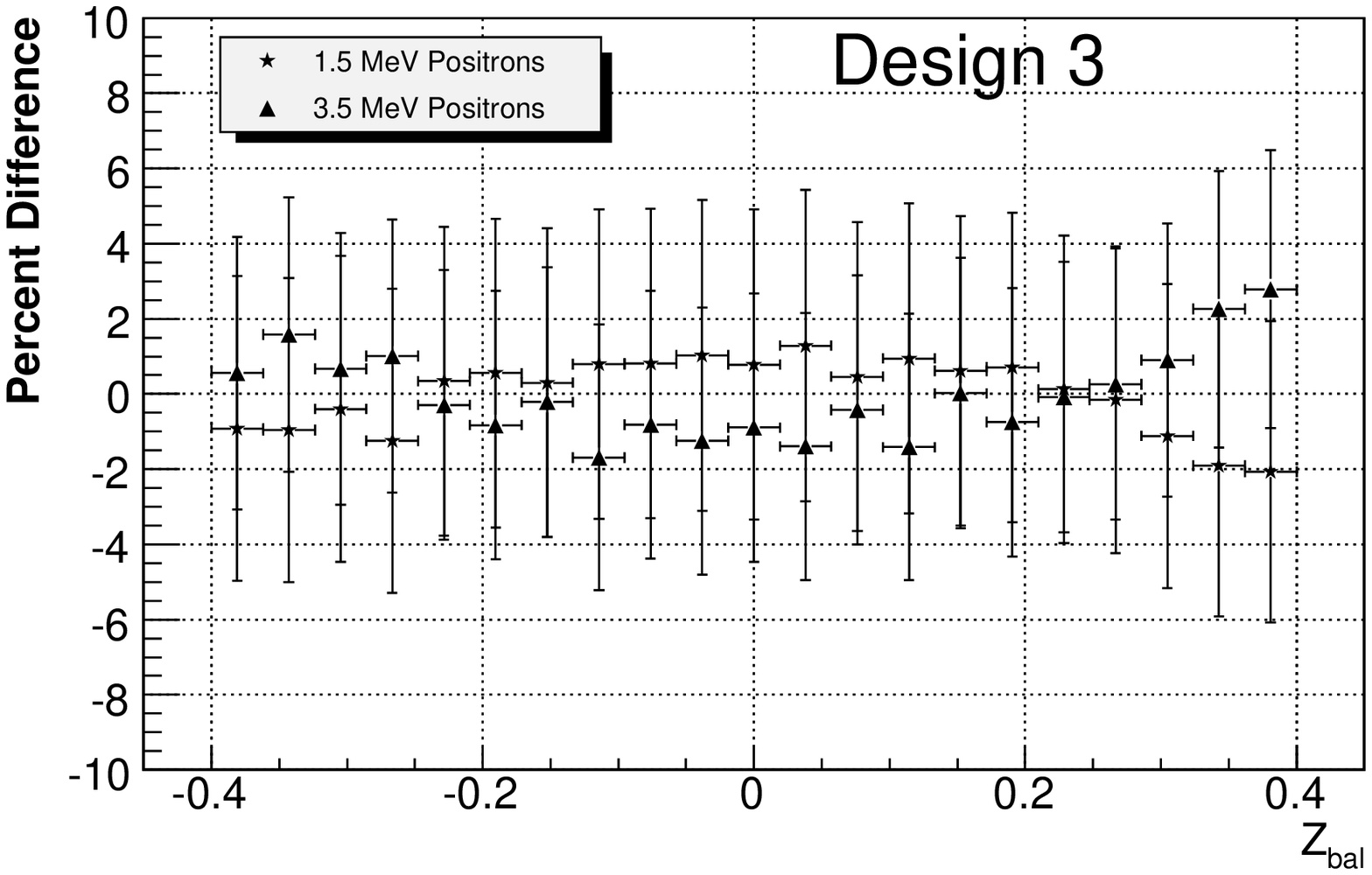}
\caption{Relative positron energy response as a function of the z-balance
as defined in Eq.~\ref{E-zBal} for detector Designs 2 and 3 after calibrations
were applied.  Multiplicative calibration factors were derived for each bin
by averaging the responses of the 1.5 and 3.5~MeV positrons shown 
Fig.~\ref{F-posUncal}.  These calibration factors were then applied 
event-by-event.  A factor of two reduction in the non-uniformity of response
was achieved for both detector designs.}
\label{F-posCal}
\end{figure}

As expected, after calibration the uniformity of the response for both 
designs is improved (See Fig.~\ref{F-posCal}).  Especially in the region 
of $Z_{bal}$ between
-.3 and .3, the relative differences are kept within $\pm$1\%---consistent
with the uncalibrated performance demonstrated in Design~1.  One can also
note the larger error for Design~3 which arises from the large response 
variation with radial position. 

\section{Conclusion}
This work was an attempt to better understand the design features
that would affect the response of a compact liquid scintillator detector to
anti-neutrinos from nuclear power reactors.  There are some interesting
trade-offs that can be identified.  For example, Designs~2 and 4 offer
an interesting compromise in efficiency and uniformity while Design~1 shows 
that very good uniformity of response can be achieved at the cost of a larger
total detector size and an efficiency for event selection of only 50\%.
On the other hand, Design~3 demonstrates the maximum effective fiducial
volume for the minimum total detector size at the cost of energy 
distortions as large as 40\%.  The optimal design will depend on the 
goals and restrictions of the specific installation.  

Clearly improvement can be made to any and all of these designs.  It was
perhaps interesting to note that the expected uniformity from 
the two-sided designs was hampered by the large radial variations in 
energy response.  Some more clever design of the PMT distribution might
be able to improve this.  Similarly, one might consider some judiciously
placed additional PMTs in Design~4 to allow for some minimal reconstruction
and calibration.  

In general, however, it appears that a compact neutrino detector with 
overall dimensions
on the order of 3~m can be built with reasonably good performance.  
Given that the energy resolution of the scintillator is usually 7--10\%,
the 15\% variation in energy response is not extreme.  It is the hope of the 
authors that this work will prove useful to those working on nuclear 
reactor monitoring and can provide fruitful directions for future detector 
development.

\end{document}